\documentclass[fleqn,twoside]{article}      
\usepackage{epsf,multicol,ifthen}
\usepackage{ujp}
\usepackage{amstext}
\usepackage{amssymb}
\usepackage{latexsym}
\usepackage{amsmath}
\usepackage{amssymb}
\numberwithin{equation}{section}

\def\sscr{\scriptscriptstyle}

\nazva{%
SELF-CONSISTENT RENORMALIZATION AS AN EFFICIENT REALIZATION
OF MAIN IDEAS OF\\
THE BOGOLIUBOV-PARASIUK R-OPERATION}%
\udk{539.12.01, 530.145}%
\nazvacol{%
SELF-CONSISTENT RENORMALIZATION: IDEAS, PROPERTIES AND THREE ALGORITHMS}%
\avtor{V.I. KUCHERYAVY}%
\avtorcol{V.I. KUCHERYAVY}%
\inst{%
Bogolyubov Institute for Theoretical Physics, Nat. Acad. Sci. of Ukraine}%
\adr{%
(14b, Metrolohichna Str., Kyiv 03680, Ukraine;
e-mail: vnkucher@bitp.kiev.ua)}%

\begin{document}
\setcounter{page}{1}%
\maketitl                 
\begin{multicols}{2}
\anot{%
Il libro della natura {\'e} scritto in lingua matematica.
\\ 
{\it Galileo Galilei,}
[\,{\it Il Saggiatore, 1623}\,].
\\[3pt]
``...At the present time, the intimate connection between causality and
the analytic continuation is revealed.
So, it is not improbable to develop of a subtraction procedure even in
the most general case by the use of analytic continuation techniques.''
\\ 
{\it O.S.\:Parasiuk,}
[\,\cite{Parasiuk1958}, {\it p.566, the last paragraph, 1956}\,].
\\[3pt]
This possibility is realized explicitly and efficiently
in a body of
our self-consistent renormalization (SCR).
By the self-consistency
is meant that all formal relations between UV-divergent Feynman amplitudes
are automatically retained as well between their regular values
obtained in the framework of the SCR.
Self-consistent renormalization
is efficiently applicable on equal grounds both to
renormalizable and nonrenormalizable theories.
SCR furnishes new means
for the constructive treatment of new subjects:
i) UV-divergence problems associated with symmetries,
Ward identities, and quantum anomalies;
ii) new relations between
finite bare and finite physical parameters of
quantum field theories.
The aim of this report
is briefly to review main ideas and properties of the SCR
and clearly to describe three mutually complementary algorithms
of the SCR that are presented
in the form maximally suited for practical applications.
}

\section{Introduction}

The keystone idea on a purely mathematical genesis of the ultraviolet (UV)
divergencies of the Feynman amplitudes (FAs) in quantum field theories
is at the heart of the Bogoliubov-Parasiuk R-operation~\cite{
Bogoliubov&Parasiuk1955,
Bogoliubov&Parasiuk1955a,
Parasiuk1955,
Bogoliubov&Parasiuk1956,
Parasiuk1956,
Bogoliubov&Parasiuk1957,
Parasiuk1958}.
Using this idea
along with related considerations of mathematicians
of the 19th and 20th centuries,\!\!~\footnote{%
It is appropriate to pointed out here that the first
regularization recipe of infinities subtraction
for turning divergent integral into convergent one had been used
in Cauchy's {\it ``extraordinary integral''}~\cite{%
Cauchy1826,
Cauchy1827,
Cauchy1844},
as well as in d'Adh{\'e}mar's~\cite{%
d'Adhemar1904,
d'Adhemar1906}
and Hadamard's~\cite{%
Hadamard1903,
Hadamard1904,
Hadamard1905,
Hadamard1908}
{\it ``finite part of divergent integral''}.
These recipes are similar but not identical.
But in both cases it was extendeed the validity of the usual rules
of change of variable, integration by parts, and differentiation with
respect to the upper limit of integration to these new objects.
The Cauchy's ``extraordinary integral''
has been used for an efficient analytic continuation of
the $\Gamma(z)$-function
to some noninteger real values ${\rm Re}\,z < 0$
firstly
by himself Cauchy~\cite{%
Cauchy1827} in 1827,
and then in the strips $(-n-1< {\rm Re}\,z < -n)$
by Saalsch\"{u}tz~\cite{%
Saalschutz1887,
Saalschutz1888}
in 1887-1888.
The term
``finite part of divergent integral''
was introduced by d'Adh{\'e}mar in his thesis presented at the Sorbonne
in December 1903, and defended in April 1904, see
[\,\cite{%
Mazya&Shaposhnikova1998}, p.477\,].
Referring to Hadamard's
article~\cite{
Hadamard1903},
d'Adh{\'e}mar
[\,\cite{d'Adhemar1906}, p.371\,],
writes ``...Independently of each other, we understood
the role of these {\it finite parts}...''.
In d'Adh{\'e}mar's thesis and articles this notion was applied to the
construction of solutions of equation for cylindrical waves~\cite{
d'Adhemar1904,
d'Adhemar1906},
whereas Hadamard use finite parts for the solution of the Cauchy problem
for second order equations with variable coefficients~\cite{
Hadamard1903,
Hadamard1904,
Hadamard1905}
and an arbitrary number of independent variables~\cite{
Hadamard1908}.
On the applications of d'Adhemar's and Hadamard's
``finite part of divergent integral''
in greater detail see Hadamard's book~\cite{
Hadamard1932}.
After 40 years later,
when analysing the connections between the intuitive and logical
ways of mathematical inventions, Hadamard~\cite{%
Hadamard1945}
wrote:
``...All mathematicians must consider themselves as logics.
For example, I have been asked by what kind of guessing I thought of
the device of the ``finite part of divergent integral'',
which I have used for the integration of partial differential equations.
Certainly, considering in itself, it lookes typically like
``thinking aside''.
But, in fact, for a long while my mind refused to conceive that idea
until positively compelled to.
I was led to it step by step as the mathematical reader will easily
verify if he takes the trouble to consult my researches on the subject,
especially my
{\it Recherches sur les solution fondamentales et l'int{'e}gration
des {'e}quations lin{'e}aires aux d{'e}riv{'e}es partielles},
2nd Memoir, especially p.121 and so on
({\it Annales Scientifiques de l'{'E}cole Normale Sup{'e}rieure,}
Vol.{\bf XXII}, 1905)~\cite{
Hadamard1905}.
I could not avoid it any more than the prisoner in Poa's tale
{\it The Pit and Pendulum}
could avoid the hole at the center of his cell...'',
see
[\,\cite{Hadamard1945}, p.110,
and p.104 or p.86 in two identical Russian translations from
French edition of 1959\,].
About further developments see M.~Riesz~\cite{%
Riesz1949,
Riesz1961},
F.~Bureau~\cite{%
Bureau1955},
R.~Courant~\cite{%
Courant1962},
and
S.G.~Samko, A.A.~Kilbas, and O.I.~Marichev~\cite{%
Samko&Kilbas&Marichev1987}.
}\; 
%
the author has developed an universal, high-efficient,
and self-consistent renormalization (SCR) technique
which is applicable for any dimension
$n=2r_n+\delta_n, \delta_n=0,1$,
$r_n\in \{0\cup\mathbb{N}_{\sscr{+}}\}$
of a space-time that is
endowed by the pseudo-euclidean $(p,q)$ metric $g^{\mu\nu}$,
where $p+q=n$, and for an arbitrary topology of Feynman graphs.

Algorithmically,
the SCR is an efficient realization of the Bogoliubov-Parasiuk R-operation
as some special analytical extension of the UV-divergent FAs
in two parameters
$\omega^{\scriptscriptstyle G}$ and
$\nu^{\mspace{1.0mu}\scriptscriptstyle G}$
by means of recurrence, compatibility, and differential
relations fixing a renormalization arbitrariness of the R-operation
in some universal way based on the mathematical properties of FAs only.
The parameters
$\omega^{\scriptscriptstyle G}$ and
$\nu^{\mspace{1.0mu}\scriptscriptstyle G}$
are depended on a space-time dimension $n$,
a graph-topological invariant $|{\cal C}|$ determining a number
of independent circuits of a graph $G$,
and two FAs characteristics
$\lambda^{\scriptscriptstyle G}$ and
$d^{\mspace{1.0mu}\scriptscriptstyle G}$.
The numbers
$\lambda^{\scriptscriptstyle G}$ and
$d^{\mspace{1.0mu}\scriptscriptstyle G}$
determine the maximal degree of polynomials
of the denominator and the numerator respectively in the integrand.
As a result, the SCR is efficiently applicable on equal grounds both to
renormalizable and nonrenormalizable theories
that is very important for quantum gravity.

By the self-consistency is meant that all formal relations
between UV-divergent FAs
are automatically retained as well between their regular values
obtained in the framework of the SCR.
The SCR furnishes new means for constructive treatment of new subjects:
i) UV-divergence problems associated with symmetries,
Ward identities, reduction identities, and quantum anomalies;
ii) new relations between
{\it finite bare} and {\it finite physical} parameters of
quantum field theories.

The aim of this report
is briefly to review main ideas and properties of the SCR, see Sect.2-3,
and clearly to describe three mutually complementary algorithms
of the SCR, see Sect.3-5,
which are presented in the form maximally suited for practical applications.

\section{The bases and possibilities of the SCR}

\noindent
{\bf 2.1}
The SCR is
an efficient realization of the Bogoliubov-Parasiuk
R-operation~\cite{
Bogoliubov&Parasiuk1955,
Bogoliubov&Parasiuk1955a,
Parasiuk1955,
Bogoliubov&Parasiuk1956,
Parasiuk1956,
Bogoliubov&Parasiuk1957,
Parasiuk1958,
Parasiuk1960}
which is supplemented with
{\it recurrence, compatibility, and differential}
relations {\it fixing a renormalization arbitrariness of the
R-operation} in some universal way based
on {\it mathematical properties} of Feynman amplitudes (FAs) only.
In its turn, the BP-approach is rested on an idea
that the {\it nature of UV-divergences is purely mathematical} and,
per se, the R-operation is a constructive form of the Hahn-Banach
theorem on extensions of linear functionals, see for example~\cite{
Stepanov1963,
Stepanov1965,
Hepp1966}.

{\bf 2.2}
Elaborating this idea the author~\cite{
Kucheryavy1974,
Kucheryavy1977,
Kucheryavy1979,
Kucheryavy1982,
Kucheryavy1982a,
Kucheryavy1983,
Kucheryavy1983a,
Kucheryavy1987,
Kucheryavy1991,
Kucheryavy1991a,
Kucheryavy1991b,
Kucheryavy1997,
Kucheryavy1997a,
Kucheryavy2002,
Kucheryavy2004}
has obtained the high-efficient realization of this renormscheme.
In this realization:

$\bullet$
Properties of special functions of the hypergeometric type
are used essentially.\!\!~\footnote{%
The connection of particular FAs with the hypergeometric functions
are well known. See, for example, investigations on analytic properties
of convergent scalar FAs by using of algebraic topology methods~\cite{%
Regge1967,
Enolsky&Golubeva1975,
Golubeva1976}, %
or calculations for needs of phenomenological physics
some classes of FAs by using of differential equation method~\cite{%
Kotikov1991,
Kotikov2001,
Argeri&Mastrolia2007}.  %
But in our case this connection is established for general divergent FAs
in any spase-time dimension $n$ and the $(p,q)$ pseudo-euqlidean metric,
$p+q=n$.
Apart from, this connection suggests some simple method of fixing
a renormalization arbitrariness of the Bogoliubov-Parasiuk $R$-operation
in some universal way based on the mathematical properties of FAs only.
As a result we obtain the self-consistent renormalization
with new valuable properties and possibilities.
}

$\bullet$
Combinatoric is simplified considerably.
Our investigations
confirm the very important assertion by D.A.Slavnov~\cite{
Slavnov1973}
that combinatorics of the R-operation is overcomplicated
considerably and can be simplified essentially.

$\bullet$
Renormalization arbitrariness of the
R-operation is fixed in such a way that basic functions
$(R_{\mspace{1.0mu}\scriptscriptstyle 0}^{\mspace{1.0mu}\nu}
{\cal F})_{sj}\equiv$
$(R_{\mspace{1.0mu}\scriptscriptstyle 0}^{\mspace{1.0mu}\nu}
{\cal F})_{sj}(\omega;M_{\epsilon},A)$
of renormalized FAs obey {\it the same recurrence relations}
as basic functions
${\cal F}_{sj}\equiv{\cal F}_{sj}(\omega;M_{\epsilon},A)$
of convergent or dimensionally regularized FAs:
\begin{gather}
M_{\epsilon}\,
{\cal F}_{s-2,j-1}-
A\,
{\cal F}_{s,j-1}+
(\omega+j)\,
{\cal F}_{sj}=0,
\notag\\ 
M_{\epsilon}\,
(R_{\mspace{1.0mu}\scriptscriptstyle 0}^{\mspace{1.0mu}\nu}
{\cal F})_{s-2,j-1}-
A\,
(R_{\mspace{1.0mu}\scriptscriptstyle 0}^{\mspace{1.0mu}\nu}
{\cal F})_{s,j-1}+
\notag\\ 
\hspace{30mm}
+\,(\omega+j)\,
(R_{\mspace{1.0mu}\scriptscriptstyle 0}^{\mspace{1.0mu}\nu}
{\cal F})_{sj}=0.
\tag{$2.1$}
\end{gather}
The explicit form of
${\cal F}_{sj}$ and
$(R_{\mspace{1.0mu}\scriptscriptstyle 0}^{\mspace{1.0mu}\nu}
{\cal F})_{sj}$
are given below by Eqs.(3.30)-(3.31).
On the self-consistent version of the Clifford aspect of the dimensional
regularization which efficiently overcomes the known difficulties
connected with $n$-dimensional generalization of the Dirac $\gamma^5$
matrix see~\cite{
Kucheryavy1992,
Kucheryavy1993}.

$\bullet$
The {\it compatibility relations} of {\it the first kind}:
\begin{align}
&
(R_{\mspace{1.0mu}\scriptscriptstyle 0}^{\mspace{1.0mu}\nu}
{\cal F})_{sj}=
{\cal F}_{sj},\
\text{ if $\nu_{sj}\!:=[(\nu-s)/2]+j\le -1$},\notag\\
&
(R_{\mspace{1.0mu}\scriptscriptstyle 0}^{\mspace{1.0mu}\nu+1}\!
{\cal F})_{s+1,j}=
(R_{\mspace{1.0mu}\scriptscriptstyle 0}^{\mspace{1.0mu}\nu}
{\cal F})_{sj},
\tag{$2.2$}
\end{align}
and the {\it compatibility relations} of {\it the second kind}:
\begin{gather}
{\cal F}_{s-2,j-1}(\omega;M_{\epsilon},A)=
{\cal F}_{s,j-1}(\omega;M_{\epsilon},A)=
\notag\\ 
\hspace{30mm}
=
{\cal F}_{sj}(\omega-1;M_{\epsilon},A),
\tag{$2.3$}\\ 
(R_{\mspace{1.0mu}\scriptscriptstyle 0}^{\mspace{1.0mu}\nu}
{\cal F})_{s-2,j-1}(\omega;M_{\epsilon},A)=
(R_{\mspace{1.0mu}\scriptscriptstyle 0}^{\mspace{1.0mu}\nu}
{\cal F})_{sj}(\omega-1;M_{\epsilon},A),
\notag\\ 
(R_{\mspace{1.0mu}\scriptscriptstyle 0}^{\mspace{1.0mu}\nu}
{\cal F})_{s,j-1}(\omega;M_{\epsilon},A)=
(R_{\mspace{1.0mu}\scriptscriptstyle 0}^{\mspace{1.0mu}\nu-2}
{\cal F})_{sj}(\omega-1;M_{\epsilon},A),
\notag
\end{gather}
are satisfied {\it automatically}.
From the first one of Eqs.{($2.2$)} it follows that formulae
for the {\it regular values obtained in the framework of SCR
describe uniformly both divergent and convergent} FAs.

$\bullet$
The {\it differential relations} for
${\cal F}_{sj}$
and
$(R_{\mspace{1.0mu}\scriptscriptstyle 0}^{\mspace{1.0mu}\nu}
{\cal F})_{sj}$
with respect to {\it mass-damping variables}
$\mu_l\!:=(m_l^2-i\epsilon_l),\ l\in {\cal L}$,
\begin{gather}
\dfrac{\partial^m}{\partial\mu_{l_1}\cdots\partial\mu_{l_m}}
\begin{bmatrix}
{\cal F}_{sj}(\omega) \\
(R_{\mspace{1.0mu}\scriptscriptstyle 0}^{\mspace{1.0mu}\nu}
{\cal F})_{sj}(\omega)
\end{bmatrix}=
\notag\\
\hspace{8mm}
=
(-1)^m\alpha_{l_1}\cdots\alpha_{l_m}
\begin{bmatrix}
{\cal F}_{sj}(\omega-m) \\
(R_{\mspace{1.0mu}\scriptscriptstyle 0}^{\mspace{1.0mu}\nu}
{\cal F})_{sj}(\omega-m)
\end{bmatrix}
\tag{$2.4$}
\end{gather}
{\it are the same},
and the {\it differential relations} for ones with respect to
{\it external momenta} $k_e,\ e\in {\cal E}$,
\begin{gather}
\partial_{\,e_1}^{\,\sigma_1}\cdots\partial_{\,\,e_m}^{\,\sigma_m}
\begin{bmatrix}
{\cal F}_{sj}(\omega) \\
(R_{\mspace{1.0mu}\scriptscriptstyle 0}^{\mspace{1.0mu}\nu}
{\cal F})_{sj}(\omega)
\end{bmatrix}
=2^{m}\sum\limits_{\varkappa=0}^{[m/2]}
{\cal A}_{\,e_1\,\cdots \,\,e_m}^{\sigma_1\cdots\,\sigma_m}
(\varkappa)
\notag\\
\hspace{8mm}
\cdot
\begin{bmatrix}
{\cal F}_{sj}(\omega-m+\varkappa) \\
(R_{\mspace{1.0mu}\scriptscriptstyle 0}^{\mspace{1.0mu}\nu-2m+2\varkappa}
{\cal F})_{sj}(\omega-m+\varkappa)
\end{bmatrix}
\tag{$2.5$}
\end{gather}
{\it are almost the same}.
Here
$\partial_{\,e_i}^{\,\sigma_i}\equiv
{\partial}/{\partial (k_{e_i})_{\sigma_i}}$,
and
${\cal A}_{\,e_1\,\cdots \,\,e_m}^{\sigma_1\cdots\,\sigma_m}(\varkappa)
\equiv
{\cal A}_{\,e_1\,\cdots \,\,e_m}^{\sigma_1\cdots\,\sigma_m}
(\varkappa|\alpha,k)$
are special homogeneous polynomials of degree $m-2\varkappa$ in
$A_{\,e_i}^{\sigma_i}\equiv A_{\,e_i}^{\sigma_i}(\alpha,k)\!:=
\sum_{e\in {\cal E}}A_{e_i e}(\alpha)k_{e}^{\sigma_i}$
and of degree $\varkappa$ in
$(\,{}_{\,e_i\,e_j}^{\sigma_i\sigma_j})\!:=
A_{e_i e_j}(\alpha)g^{\sigma_i\sigma_j}$
where
$A_{ee'}(\alpha)$
are matrix elements of the quad\-ratic Kirchhoff form in external
momenta $k_e,\ e\in {\cal E}$.
The polynomials
${\cal A}_{\,e_1\,\cdots \,\,e_m}^{\sigma_1\cdots\,\sigma_m}
(\varkappa|\alpha,k)$
have an algebraic structure of quantities generated by the Wick
formula, which represents a $T$-product of $m$ boson fields in terms
of some set of $N$-products of $m-2\varkappa$ boson fields with
$\varkappa$ primitive contractions.
Here the quantities
$A_{\,e_i}^{\sigma_i}$ and
$(\,{}_{\,e_i\,e_j}^{\sigma_i\sigma_j})$
play the role of boson fields and their contractions respectively.

$\bullet$
It is essential that
${\cal F}_{sj}$ and
$(R_{\mspace{1.0mu}\scriptscriptstyle 0}^{\mspace{1.0mu}\nu}
{\cal F})_{sj}$
as {\it functions of two variables} $M_{\epsilon}$ and $A$ are
the {\it homogeneous functions of the same degree} $\omega +j$.
From this it follows that they \linebreak %
are solutions to the same partial differential equations,
namely to the {\it Euler equation} for homogeneous functions
\begin{equation}
\bigl[
M_{\epsilon}
\partial_{\scriptscriptstyle M_{\epsilon}}+
A
\partial_{\scriptscriptstyle A} - (\omega+j)
\bigr]
\begin{bmatrix}
{\cal F}_{sj}(\omega) \\
\,(R_{\mspace{1.0mu}\scriptscriptstyle 0}^{\mspace{1.0mu}\nu}
{\cal F})_{sj}(\omega)\,
\end{bmatrix}=0,
\tag{$2.6$}
\end{equation}
and to some family of second order equations emerging from
Eq.{($2.6$)},
for example
\begin{gather}
\bigl[
M_{\epsilon}
\partial^2_{\scriptscriptstyle M_{\epsilon}M_{\epsilon}}
\pm(M_{\epsilon} \pm A)
\partial^2_{\scriptscriptstyle M_{\epsilon}A}
\pm A
\partial^2_{\scriptscriptstyle AA}-
\notag\\
-(\omega+j-1)
({\partial}_{\scriptscriptstyle M_{\epsilon}}
\pm
{\partial}_{\scriptscriptstyle A} )
\bigr]
\begin{bmatrix}
{\cal F}_{sj}(\omega) \\[2pt]
\,(R_{\mspace{1.0mu}\scriptscriptstyle 0}^{\mspace{1.0mu}\nu}
{\cal F})_{sj}(\omega)\,
\end{bmatrix}=0,
\tag{$2.7$}
\end{gather}
that can be again represented as the Euler equation
\begin{gather}
\bigl[
M_{\epsilon}
\partial_{\scriptscriptstyle M_{\epsilon}}+
A
\partial_{\scriptscriptstyle A}-
(\omega+j-1)
\bigr]
\cdot
\notag\\
\hspace{15mm}
\cdot
\begin{bmatrix}
({\partial}_{\scriptscriptstyle M_{\epsilon}}\pm
{\partial}_{\scriptscriptstyle A})
{\cal F}_{sj}(\omega) \\
({\partial}_{\scriptscriptstyle M_{\epsilon}}\pm
{\partial}_{\scriptscriptstyle A})
\,(R_{\mspace{1.0mu}\scriptscriptstyle 0}^{\mspace{1.0mu}\nu}
{\cal F})_{sj}(\omega)\,
\end{bmatrix}=0.
\tag{$2.8$}
\end{gather}
So,
an important role of the quantities
$({\partial}_{\scriptscriptstyle M_{\epsilon}}\pm
{\partial}_{\scriptscriptstyle A})
{\cal F}_{sj}(\omega)$
and
$({\partial}_{\scriptscriptstyle M_{\epsilon}}\pm
{\partial}_{\scriptscriptstyle A})
(R_{\mspace{1.0mu}\scriptscriptstyle 0}^{\mspace{1.0mu}\nu}
{\cal F})_{sj}(\omega)$
is revealed in our problem.
After repeating this procedure ${\scriptstyle N}+1$ times
one obtains
\begin{gather}
\bigl[
M_{\epsilon}
\partial_{\scriptscriptstyle M_{\epsilon}}+
A
\partial_{\scriptscriptstyle A}
-(\omega+j-{\scriptstyle N}-1)
\bigr]
\cdot
\notag\\
\hspace{10mm}
\cdot
\begin{bmatrix}
({\partial}_{\scriptscriptstyle M_{\epsilon}}\pm
{\partial}_{\scriptscriptstyle A})
{\cal F}_{sj}^{\scriptscriptstyle N_{\pm}}
(\omega-{\scriptstyle N}) \\
({\partial}_{\scriptscriptstyle M_{\epsilon}}\pm
{\partial}_{\scriptscriptstyle A})
\,(R_{\mspace{1.0mu}\scriptscriptstyle 0}^{\mspace{1.0mu}\nu}
{\cal F})_{sj}^{\scriptscriptstyle N_{\pm}}
(\omega-{\scriptstyle N})\,
\end{bmatrix}=0,
\tag{$2.9$}
\end{gather}
where we define
${\cal F}_{sj}^{\scriptscriptstyle N_{\pm}}
(\omega-{\scriptstyle N})\!:=$
$({\partial}_{\scriptscriptstyle M_{\epsilon}}\pm
{\partial}_{\scriptscriptstyle A})^{\scriptscriptstyle N}
{\cal F}_{sj}(\omega)$
and
$(R_{\mspace{1.0mu}\scriptscriptstyle 0}^{\mspace{1.0mu}\nu}
{\cal F})_{sj}^{\scriptscriptstyle N_{\pm}}
(\omega-{\scriptstyle N})\!:=$
$({\partial}_{\scriptscriptstyle M_{\epsilon}}\pm
{\partial}_{\scriptscriptstyle A})^{\scriptscriptstyle N}
(R_{\mspace{1.0mu}\scriptscriptstyle 0}^{\mspace{1.0mu}\nu}
{\cal F})_{sj}(\omega)$.
If ${\scriptstyle N}$ such that $(\omega-{\scriptstyle N}+j)\le -\,1$
then both
$({\partial}_{\scriptscriptstyle M_{\epsilon}}+
{\partial}_{\scriptscriptstyle A})
{\cal F}_{sj}^{\scriptscriptstyle N_{\pm}}
(\omega-{\scriptstyle N})=0$
and
$({\partial}_{\scriptscriptstyle M_{\epsilon}}+
{\partial}_{\scriptscriptstyle A})
(R_{\mspace{1.0mu}\scriptscriptstyle 0}^{\mspace{1.0mu}\nu}
{\cal F})_{sj}^{\scriptscriptstyle N_{\pm}}
(\omega-{\scriptstyle N})=0$.
As a result, Eq.(2.9) with the {\it plus sign} is degenerated
into the {\it identical zero} and with the {\it minus sign} is reduced
to the {\it Euler--Poisson--Darboux equation}
\begin{gather}
\biggl[
\frac {\partial^2} {\partial M_{\epsilon}\partial A}+
\frac {(\omega+j-{\scriptstyle N}-1)/2} {M_{\epsilon}-A}
\Bigl(
\frac {\partial}{\partial M_{\epsilon}}-
\frac {\partial}{\partial A}
\Bigr)
\biggr]
\cdot
\notag\\
\hspace{20mm}
\cdot
\begin{bmatrix}
{\cal F}_{sj}^{\scriptscriptstyle N_{\pm}}
(\omega-{\scriptstyle N}) \\[2pt]
\,(R_{\mspace{1.0mu}\scriptscriptstyle 0}^{\mspace{1.0mu}\nu}
{\cal F})_{sj}^{\scriptscriptstyle N_{\pm}}
(\omega-{\scriptstyle N})\,
\end{bmatrix}=0.
\tag{$2.10$}
\end{gather}
Consistency of solutions to Eqs.(2.9)-(2.10)
for different preassigned asymptotic of
$(R_{\mspace{1.0mu}\scriptscriptstyle 0}^{\mspace{1.0mu}\nu}
{\cal F})_{sj}$
at the vicinity of $A=0$
leads to relations
\begin{gather}
{\partial}_{\scriptscriptstyle M_{\epsilon}}
{\cal F}_{sj}(\omega)=
-{\cal F}_{sj}(\omega-1),
\notag\\[2pt]
{\partial}_{\scriptscriptstyle A}
{\cal F}_{sj}(\omega)=
{\cal F}_{sj}(\omega-1),
\notag\\[2pt]
{\partial}_{\scriptscriptstyle M_{\epsilon}}
(R_{\mspace{1.0mu}\scriptscriptstyle 0}^{\mspace{1.0mu}\nu}
{\cal F})_{sj}(\omega)=
-(R_{\mspace{1.0mu}\scriptscriptstyle 0}^{\mspace{1.0mu}\nu}
{\cal F})_{sj}(\omega-1),
\tag{$2.11$}\\[2pt]
{\partial}_{\scriptscriptstyle A}
(R_{\mspace{1.0mu}\scriptscriptstyle 0}^{\mspace{1.0mu}\nu}
{\cal F})_{sj}(\omega)=
(R_{\mspace{1.0mu}\scriptscriptstyle 0}^{\mspace{1.0mu}\nu-2}
{\cal F})_{sj}(\omega-1),
\notag
\end{gather}
that are also followed from the explicit form of the basic functions
${\cal F}_{sj}$ and
$(R_{\mspace{1.0mu}\scriptscriptstyle 0}^{\mspace{1.0mu}\nu}
{\cal F})_{sj}$,
see Eqs.(3.30)-(3.31) below.

{\bf 2.3}
Relations (2.1)-(2.11) manifest a mutual consistency of
asymptotic properties
of different terms of FAs with respect to external momenta and masses.
It is precisely these
{\it recurrence, compatibility, and differential relations}
that are of great importance for
investigating of {\it symmetries and anomalies problems}
and for turning  of developed renormscheme
into {\it the self-consis\-tent one}.

Besides,
there exist some obvious identities
of generic nature which are called as the
{\it reduction identities} (RIs)~\cite{ 
Kucheryavy1991a,
Kucheryavy1991b}
that in another way lead to the recurrence relations (2.1).
The simple idea of cancelling of equal factors
in factorized polynomials
in a numerator and a denominator of integrands is used in RIs.
The reduction identities also are of
great importance for applications
as {\it an origin new nontrivial identities}.
Some of them have been used essentially in our investigations~\cite{%
Kucheryavy1991,
Kucheryavy1991a,
Kucheryavy1991b,
Kucheryavy1997a,
Kucheryavy2002,
Kucheryavy2004,
Kucheryavy2000,
Kucheryavy2001,
Kucheryavy2002a,
Kucheryavy2004a}.

{\bf 2.4}
From Eqs.(2.1)-(2.11) and the explicit form of the basic functions
${\cal F}_{sj}$,\
$(R_{\mspace{1.0mu}\scriptscriptstyle 0}^{\mspace{1.0mu}\nu}
{\cal F})_{sj}$,
see Eqs.(3.30)-(3.33) and (3.36)-(3.40),
implies the following important properties of the SCR:

\noindent
{\bf Algorithmic universality.}
The SCR is a special analytic continuation of any FA
firstly given by an UV-divergent integral.
In so doing divergence indices
$\nu$ of FA's may be as large as one needs.
Hereafter,
this continuation will be named
as the regular (i.e., finite) value of this FA.
As a result, the regular values of FA's
respect certain
recurrence, compatibility, and differential properties of
an universal character and
have already been realized efficiently as convergent integrals.
Therefore,
the calculation of FA's corresponding to renormalizable and
nonrenormalizable theories does not differ for the two in the
framework of this renormscheme.
Actually, the problem is reduced to calculations of
the characteristic numbers,
$\omega$, $\nu_{sj}$, and
$\lambda_{sj}$ determining the basic functions
$(R_{\mspace{1.0mu}\scriptscriptstyle 0}^{\mspace{1.0mu}\nu}
{\cal F})_{sj}$.
\smallskip

\noindent
{\bf Separation of problems. }
The SCR clearly and efficiently separate the problem of evaluating
regular values of UV-divergent quantities of
quantum field theories from that of
relations between bare and physical parameters of these theories,
i.e., the SCR realizes in practice this the very important potential
possibility of the Bogoliubov-Parasiuk R-operation.
\smallskip

\noindent
{\bf Conservation of relations.}
Any formal relation between UV-divergent quantities will be retained
also between regular values of those if the regular
values of all quantities involved in this relation
are calculated by {\it the same renormalization index} $\nu$
(the maximum one since, otherwise, we cannot guarantee
the finiteness of all terms in the relation).
So, the SCR is automatically consistent with the correspondence
principle.
As a result, the regular values obtained in the framefork of the SCR
do satisfy the vector and axial-vector
canonical Ward identities (CWIs) simultaneously.
\smallskip

\noindent
{\bf Extraction of anomalies (quantum corrections).}
In the SCR, owing to the analytic continuation technique,
quantum anomalies
(i.e., quantum corrections (QCs) more exactly)
are automatically accounted for in quantities satisfying the CWIs.
More specifically,
quantum anomalies (i.e., QCs) reveal themselves either as
an oversubtraction effect for a non-chiral case
and for the chiral limit case
(in these cases the Schwinger terms contributions (STCs)
of current commutators are zero)
or as the nonzero STCs for the chiral case.
If necessary, the explicit form of quantum anomalies (i.e., QCs)
can be easily extracted as a difference between two regular values
of the same UV-divergent quantity calculated for proper and
improper divergence indices.
\smallskip

{\bf 2.5}
Algorithmically,
the SCR is a union of three
efficient algorithms of finding:\\
\indent
i) the convergent $\alpha$-parametric integral representations of
re\-normalized FAs with a compact domain of integration of the simplex
type and with the self-consistent basic functions
$(R_{\mspace{1.0mu}\scriptscriptstyle 0}^{\mspace{1.0mu}\nu}
{\cal F})_{sj}$,
$s=0,\ldots,d^{\,\scriptscriptstyle G}$, $j=0,\ldots,[s/2]$;\\
\indent
ii) the homogeneous $k$-polynomials
${\cal P}^{\,\scriptscriptstyle G}_{sj}(m,\alpha,k)$,
$j=0,1,\ldots,[s/2]$, of degree $(s-2j)$
in external momenta $k_e,\, e\in {\cal E}$,
being as $\alpha$-parametric images of homogeneous $p$-polynomials
${\cal P}^{\,\scriptscriptstyle G}_s(m,p)$,
$s=0,\ldots,d^{\,\scriptscriptstyle G}$, of degree $s$
in internal momenta $p_l,\, l\in {\cal L}$;\\
\indent
iii) the $\alpha$-parametric functions
$\Delta(\alpha)$, $A(\alpha,k)$, $Y_l(\alpha,k)$, $X_{ll'}(\alpha)$,
$l,l'\in {\cal L}$.

\section{Parametric integral representations and basic functions
of FAs in the SCR}

\noindent
{\bf 3.1}
From the mathematical point of view any Feynman amplitude
(FA) associated with an oriented graph $G$,
\begin{equation*}
G\!:=<{\cal V},{\cal L}\cup{\cal E}\mid e_{il}=0,\,\pm 1,\
v_i\in {\cal V},\ l\in {\cal L}\cup{\cal E}>,
\end{equation*}
in which
${\cal V}$
is a set of vertices;\
${\cal L}$
is a set of internal lines;\
${\cal E}$
is a set of external lines;\
and
$e_{il}$
is an incidence matrix
(i.e., a vertex-line incidence matrix) such that:
$e_{il}=0$ if the line $l\in {\cal L}\cup{\cal E}$ is nonincident
to the vertex $v_i\in{\cal V}$;
$e_{il}=1$ if the line $l\in {\cal L}\cup{\cal E}$ is outgoing
from the vertex $v_i\in{\cal V}$;
$e_{il}=-1$ if the line $l\in {\cal L}\cup{\cal E}$ is incoming
to the vertex $v_i\in{\cal V}$,
can be always represented by the integral
\begin{gather}
I^{\scriptscriptstyle G}(m,k)_{\epsilon}\!:=
c^{\mspace{1.0mu}\scriptscriptstyle G}\!
\int_{-\infty}^{\infty}
(d^{\,n}p)^{\scriptscriptstyle{\cal L}}
\delta^{\mspace{1.0mu}\scriptscriptstyle G}(p,k)
\dfrac { {\cal P}^{\mspace{1.0mu}\scriptscriptstyle G}(m,p) }
{Q^{\mspace{1.0mu}\scriptscriptstyle G}(m,p)_{\epsilon}},\
\notag\\[2pt]
(d^{\,n}p)^{\scriptscriptstyle{\cal L}}:=
d^{\,n}p_1\cdots d^{\,n}p_{|\cal L|},\ \
d^{\,n}p_l\!:={\textstyle\prod\limits}_{\sigma=1}^n
d\mspace{1.0mu}p_l^\sigma,
\notag\\[2pt]
l\in {\cal L},\ \
m\!:=(m_1,\ldots, m_{|\cal L|}),
\tag{$3.1$}\\[2pt]
p\!:=(p_1,\ldots, p_{|\cal L|}), \ \
k\!:=(k_1,\ldots, k_{|\cal E|}).
\notag
\end{gather}
Here
${\cal P}^{\mspace{1.0mu}\scriptscriptstyle G}(m,p)$
and
$Q^{\mspace{1.0mu}\scriptscriptstyle G}(m,p)$
are a numerator and a denominator polynomials
\begin{gather}
{\cal P}^{\mspace{1.0mu}\scriptscriptstyle G}(m,p)\!:=
{\textstyle\prod\limits}_{v_i\in{\cal V}}\, P_i (m,p)
{\textstyle\prod\limits}_{l\in{\cal L}}\, P_l (m,p)=
\notag\\[2pt]
%
\phantom{%
{\cal P}^{\mspace{1.0mu}\scriptscriptstyle G}(m,p)\!:
}%
=
{\textstyle\sum\limits}_{s=0}^{d^{\mspace{1.0mu}\scriptscriptstyle G}}\,
{\cal P}_s^{\mspace{1.0mu}\scriptscriptstyle G}(m,p),
\tag{$3.2$}\\[2pt]
%
Q^{\mspace{1.0mu}\scriptscriptstyle G}(m,p)_{\epsilon}\!:=
{\textstyle\prod\limits}_{l\in {\cal L}}\,
(\mu_{l\epsilon}-p_l^2)^{\lambda_l},
\notag\\[2pt]
%
\mu_{l\epsilon}\!:=m_l^2-i\epsilon_l,\
m_l\ge 0,\ \epsilon_l>0,\
\lambda_l\in \mathbb{N}_{+},\
\forall l\in{\cal L},
\notag 
\end{gather}
$\delta^{\mspace{1.0mu}\scriptscriptstyle G}(p,k)$
is a product of vertex $\delta$-functions
\begin{gather}
\delta^{\mspace{1.0mu}\scriptscriptstyle G}(p,k)\!:=
{\textstyle\prod\limits}_{v_i\in{\cal V}}\,
\delta_i(p,k),
\notag\\[3pt]
\delta_i (p,k)\!:=\delta
\big(
{\textstyle\sum\limits}_{l\in{\cal L}}\,
e_{il}p_l+
{\textstyle\sum\limits}_{e\in{\cal E}}\,
e_{ie}k_e
\big);
\tag{$3.3$}
\end{gather}
$|{\cal A}|$ is a number of elements of some finite set ${\cal A}$;\
$\mathbb{N}_{+}$ is the set positive integers;\
${\cal P}_s^{\mspace{1.0mu}\scriptscriptstyle G}(m,p)$,
$s=0,\ldots,d^{\mspace{1.0mu}\scriptscriptstyle G}$,
are $s$-degree homogeneous polynomials in internal momenta
$p_l,\, l\in {\cal L}$;\
$P_i (m,p)$ and $P_l(m,p)$
are multiplicative generating polynomials of the numerator
${\cal P}^{\mspace{1.0mu}\scriptscriptstyle G}(m,p)$
that are
corresponded to the vertex $v_i$-contribution
$V_i (m,p,k)$, and to the internal line $l$-contribution
$\Delta_l (m,p)_{\epsilon}$, respectively:
\begin{gather}
V_i (m,p,k)\!:=P_i (m,p)\delta_i (p,k),
\notag\\ 
deg_p P_i (m,p) =:\!d_{\mspace{1.0mu}i}\ge 0,\ \
\forall v_i\in{\cal V},
\notag\\ 
\Delta_l (m,p)_{\epsilon}\!:=
\dfrac {P_l (m,p)} {(\mu_{l\epsilon}-p_l^2)^{\lambda_l}},
\tag{$3.4$}\\ 
deg_p P_l (m,p)=:\!d_l\ge 0,\ \ \forall l\in{\cal L}.
\notag
\end{gather}
The non-degenerate metric form
\begin{gather}
{\rm diag}\,g^{\mu\nu}\!:=(\,\underbrace{1,\ldots,1}_{p},
\underbrace{-1,\ldots,-1}_{q}\,),\ \
\tag{$3.5$}\\ 
p+q=n=2r_n+\delta_n,\ \
\delta_n=0,1,\ \
r_n\in \{0\cup\mathbb{N}_{\sscr{+}}\},
\notag
\end{gather}
is used for each $n$-dimensional $p_l$-integration in Eq.(3.1).

\smallskip
{\bf 3.2}
Two characreristics
\begin{gather}
\nu^{\mspace{1.0mu}\scriptscriptstyle G}\!:=
2\,\omega^{\mspace{1.0mu}\scriptscriptstyle G}+
d^{\mspace{1.0mu}\scriptscriptstyle G},\qquad
\omega^{\mspace{1.0mu}\scriptscriptstyle G}\!:=
({n}/{2})|{\cal C}|
- \lambda_{\scriptscriptstyle {\cal L}},
\notag\\[3pt]
|{\cal C}|=|{\cal L}|-|{\cal V}|+1, \quad
\lambda_{\scriptscriptstyle {\cal L}}\!:=
{\textstyle\sum\limits}_{l\in{\cal L}}\,\lambda_l,
\notag\\[3pt]
d^{\mspace{1.0mu}\scriptscriptstyle G}\!:=
d_{\scriptscriptstyle {\cal V}} + d_{\scriptscriptstyle {\cal L}}=
{\textstyle\sum\limits}_{v_i\in{\cal V}}\,d_{\mspace{1.0mu}i}+
{\textstyle\sum\limits}_{l\in{\cal L}}\,d_l,
\tag{$3.6$}
\end{gather}
of the integral (3.1) are especially important. Here
$|{\cal C}|$ is a number of independent circuits of the graph $G$.
There exist analogous characteristics for all
one particle irreducible (1PI) subgraphs
${\overline G} \subset G$.
If $\nu^{\mspace{1.0mu}\scriptscriptstyle G}\geq\,0$ or
$\nu^{\overline{\scriptscriptstyle G}}\geq\,0$ for some 1PI
${\overline G} \subset G$,
the integral is UV-divergent and a renormalization is needed~\cite{
Stepanov1963,
Hepp1966}.

While Eqs.(3.1)-(3.3) are identical
to the well-known representation in terms of vertex-line contributions,
\begin{equation*}
\delta^{\mspace{1.0mu}\scriptscriptstyle G}(p,k)
\frac
{{\cal P}^{\mspace{1.0mu}\scriptscriptstyle G}(m,p)}
{Q^{\mspace{1.0mu}\scriptscriptstyle G}(m,p)_{\epsilon}}=
\prod_{v_i\in {\cal V}}V_i (m,p,k)
\prod_{l\in {\cal L}}\Delta_l (m,p)_{\epsilon},
\end{equation*}
they are more suited for practical calculations.
The universal decomposition of
${\cal P}^{\mspace{1.0mu}\scriptscriptstyle G}(m,p)$
in terms of
$s$-degree homogeneous $p$-polynomials
${\cal P}_s^{\mspace{1.0mu}\scriptscriptstyle G}(m,p)$
is very useful.

\smallskip
{\bf 3.3}
We shall use of the Fock--Schwinger exponential $\alpha$-representation,
see for example,~\cite{%
Fock1937,
Schwinger1951,
Bogoliubov&Shirkov1957},
along with the Hepp regularization~\cite{%
Hepp1966}
that introduces parameters $r_l>0$ in the vicinity of $\alpha_l=0$,
$\forall{l}\in{\cal L}$,
\begin{gather}
\dfrac{1}
{(\mu_{l\epsilon}-p_l^2)^{\lambda_l}}=
\lim_{r_l\to 0}
\int_{r_l}^{\infty}
\dfrac
{d\alpha_l\,\alpha_l^{\lambda_l-1}i^{\lambda_l}}
{\Gamma(\lambda_l)}
e^{-i\alpha_l(\mu_{l\epsilon}-p_l^2)},
\notag\\[4pt]
p_l^{\tau}=
(-i\partial/{\partial q_{l\tau}})
e^{i(p_l\cdot q_l)}|_{q_l=0},\
\tag{$3.7$}\\[4pt]
(p_l\cdot q_l)\!:=p_{l\tau}q_{l\sigma}g^{\tau\sigma},\ \
0<r_l\leq\alpha_l\leq\infty,\
\forall{l}\in{\cal L}.
\notag
\end{gather}
Then the ratio of polynomials
${\cal P}^{\mspace{1.0mu}\scriptscriptstyle G}(m,p)/
Q^{\mspace{1.0mu}\scriptscriptstyle G}(m,p)_{\epsilon}$
in Eqs.(3.1)-(3.2) can be represented in the form
\begin{gather}
\dfrac {{\cal P}^{\mspace{1.0mu}\scriptscriptstyle G}(m,p)}
{Q^{\mspace{1.0mu}\scriptscriptstyle G}(m,p)_{\epsilon}}=
\lim_{\substack{ r_l \to 0\\ \forall{l}\in{\cal L} }}
\biggl\{%
\int_{ {R}_{+}^{\scriptscriptstyle|{\cal L}|}(\vec{r}\,) }
dv^{\scriptscriptstyle G}(\alpha)\,
i^{\lambda_{\scriptscriptstyle {\cal L}}}
\notag\\ 
\cdot
\sum_{s=0}^{d^{\mspace{1.0mu}\scriptscriptstyle G}}
{\cal P}_s^{\mspace{1.0mu}\scriptscriptstyle G}
(m,-i {\partial}/ {\partial{q_{\scriptscriptstyle{\cal L}}}})\,
e^{%
-iM_{\epsilon}%
\,+\,
iW^{q_{\scriptscriptstyle {\cal L}}}_{p_{\scriptscriptstyle {\cal L}}}
}%
\biggr\}
{\bigg|}_{%
\begin{subarray}{l}
q_l=0\\ \forall{l}\in{\cal L}
\end{subarray} },
\notag\\ 
W^{q_{\scriptscriptstyle {\cal L}}}_{p_{\scriptscriptstyle {\cal L}}}\!:=
(p^{\scriptscriptstyle T}_{\scriptscriptstyle {\cal L}}\cdot
\alpha_{\scriptscriptstyle {\cal L}{\cal L} }
p_{\scriptscriptstyle {\cal L}}) +
(p^{\scriptscriptstyle T}_{\scriptscriptstyle {\cal L}}\cdot
q_{\scriptscriptstyle {\cal L}})
\tag{$3.8$}\\[2pt]
=\sum_{l\in{\cal L}}\,\alpha_lp_l^2
+
\sum_{l\in{\cal L}}\,(p_l\cdot q_l),\ \
[\alpha_{\scriptscriptstyle {\cal L}{\cal L}}]_{ll'}\!:=
\alpha_l\,\delta_{ll'}.
\notag
\end{gather}
In Eq.(3.8) the
$p_{\scriptscriptstyle {\cal L}}$
and
$q_{\scriptscriptstyle {\cal L}}$
are $(|{\cal L}|\times n)$-dimensional actual and auxilary
internal momenta column-vectors
associated with the set of internal lines, ${\cal L}$, of a graph $G$;
the ${\scriptstyle T}$ is the transpose sign, so
$p^{\scriptscriptstyle T}_{\scriptscriptstyle {\cal L}}$
is the row-vector;
the $\alpha_{\scriptscriptstyle {\cal L}{\cal L}}$
is the $|{\cal L}|$-dimensional diagonal matrix of $\alpha$-parameters;
the $\lambda_{\scriptscriptstyle {\cal L}}$ is defined in Eq.(3.6).
Here, the integration measure
$dv^{\scriptscriptstyle G}(\alpha)$,
the integration region
${R}_{+}^{\scriptscriptstyle|{\cal L}|}(\vec{r}\,)$,
and the $\alpha$-parametric function
$M_{\epsilon}\equiv M(m,\alpha)_{\epsilon}$
which is the linear form in
the square of internal masses with $i\epsilon$-damping
are defined as,
\begin{gather}
dv^{\scriptscriptstyle G}(\alpha)\!:=
\prod_{l\in {\cal L}}
\biggl(
\frac {d\alpha_l\,\alpha_l^{\lambda_l-1}} {\Gamma(\lambda_l)}
\biggr),
\notag\\[1pt]
{R}_{+}^{\scriptscriptstyle|{\cal L}|}(\vec{r}\,)\!:=
\{
\alpha_l|\,0<r_l\leq\alpha_l\leq\infty,\
\forall{l}\in{\cal L},
\},\
\notag\\[2pt]
M_{\epsilon}\!:=\sum_{l\in{\cal L}}\,
\alpha_l\mu_{l\epsilon},\quad
\mu_{l\epsilon}\!:=(m_l^2-i\epsilon_l).
\tag{$3.9$}
\end{gather}
Now, substituting of Eq.(3.8) into Eq.(3.1) and interchanging
the order of integration in $p_l$
and $\alpha_l$, $\forall{l}\in{\cal L}$, we obtain
the very useful representation of the regularized by Hepp integral
$I^{\scriptscriptstyle G}(m,k)^{\vec{r}}_{\epsilon}$.
The integrand of it is the
$(|{\cal L}|\times n)$-dimensional pseudo-Euclidean
Gaussian-like expression but in depended variables
$p_l, \forall{l}\in{\cal L}$,
\begin{gather}
I^{\scriptscriptstyle G}(m,k)^{\vec{r}}_{\epsilon}\!:=
c^{\mspace{1.0mu}\scriptscriptstyle G}
\int_{ {R}_{+}^{\scriptscriptstyle|{\cal L}|}(\vec{r}\,) }
dv^{\scriptscriptstyle G}(\alpha)
\notag\\ 
\cdot
\sum_{s=0}^{d^{\mspace{1.0mu}\scriptscriptstyle G}}
{\cal P}_s^{\mspace{1.0mu}\scriptscriptstyle G}
(m,-i {\partial}/ {\partial{q_{\scriptscriptstyle{\cal L}}}})\,
\int\limits_{-\infty}^{\infty}
(d^{\,n}p)^{\scriptscriptstyle{\cal L}}
\delta^{\mspace{1.0mu}\scriptscriptstyle G}
(p_{\sscr{\cal L}},k_{\sscr{\cal E}})\,
i^{\lambda_{\scriptscriptstyle {\cal L}}}
\notag\\[-4pt]
\phantom{%
\cdot
\sum_{s=0}^{d^{\mspace{1.0mu}\scriptscriptstyle G}}
{\cal P}_s^{\mspace{1.0mu}\scriptscriptstyle G}
(m,-i
}%
\cdot\,
e^{%
-iM_{\epsilon}%
\,+\,
iW^{q_{\scriptscriptstyle {\cal L}}}_{p_{\scriptscriptstyle {\cal L}}}
}%
{\bigg|}_{%
\begin{subarray}{l}
q_l=0\\ \forall{l}\in{\cal L}
\end{subarray} }.
\tag{$3.10$}
\end{gather}

The set of internal lines, ${\cal L}$, can be always decompose
(as a rule in more than one way) into two mutually disjoint subsets,
${\cal L}={\cal M}\cup{\cal N},\ {\cal M}\cap{\cal N}=\emptyset$,
which determine some {\it skeleton tree},
i.e., {\it 1-tree} subgraph
$G({\cal V},{\cal M}\cup{\cal E})$, with $|{\cal M}|=|{\cal V}|-1$,
and corresponding to it {\it co-tree} subgraph
$G({\cal V},{\cal N}\cup{\cal E})$, with
$|{\cal N}|= |{\cal L}|-|{\cal V}|+1=|{\cal C}|$
of the graph $G$.
Supports of all
$\delta_i (p_{\sscr{\cal L}},k_{\sscr{\cal E}})$-functions,
$\forall{v_i}\in{\cal V}$,
see Eq.(3.3),
are defined by Eqs.(3.11) and are equivalent to the matrix relations
given in Eqs.(3.12) and Sec.5,
\begin{gather}
{\textstyle\sum\limits}_{l\in{\cal L}}\,
e_{il}p_l+
{\textstyle\sum\limits}_{e\in{\cal E}}\,
e_{ie}k_e=0,\
\forall{v_i}\in{\cal V},
\tag{$3.11$}\\[1pt]
e_{\{\sscr{\cal V}/j\}\cal M}p_{\sscr{\cal M}} +
e_{\{\sscr{\cal V}/j\}\cal N}p_{\sscr{\cal N}} +
e_{\{\sscr{\cal V}/j\}\cal E}k_{\sscr{\cal E}}=
0_{\{\sscr{\cal V}/j\}},
\notag\\[1pt]
e_{\sscr{j\cal M}}p_{\sscr{\cal M}} +
e_{\sscr{j\cal N}}p_{\sscr{\cal N}} +
e_{\sscr{j\cal E}}k_{\sscr{\cal E}}=0,\
\text{$v_j$\,--\,the basis vertex},
\notag\\[1pt]
p_{\scriptscriptstyle{\cal L}}=
e_{\sscr{\cal L}{\cal N}}p_{\sscr{\cal N}}
+
e_{\sscr{\cal L}{\cal E}}(j)k_{\sscr{\cal E}},
\notag\\[0pt]
p_{\scriptscriptstyle{\cal M}}=
e_{\sscr{\cal M}{\cal N}}p_{\sscr{\cal N}}
+
e_{\sscr{\cal M}{\cal E}}(j)k_{\sscr{\cal E}}.
\tag{$3.12$}
\end{gather}
Thus, $(|{\cal M}|\times n)$-dimensional integration by means of
$\delta_i (p_{\sscr{\cal L}},k_{\sscr{\cal E}})$-functions,
$\forall{v_i}\in{{\cal V}/j}$,
(this is equivalent to make use of Eqs.(3.12)),
gives rise to the following intermediate $\alpha$-parametric
representation,
\begin{gather}
I^{\scriptscriptstyle G}(m,k)^{\vec{r}}_{\epsilon}\!:=
\delta^{\mspace{1.0mu}\scriptscriptstyle G}(k_{\sscr{\cal E}})
c^{\mspace{1.0mu}\scriptscriptstyle G}
\int_{ {R}_{+}^{\scriptscriptstyle|{\cal L}|}(\vec{r}\,) }
dv^{\scriptscriptstyle G}(\alpha)
\notag\\ 
\cdot
\sum_{s=0}^{d^{\mspace{1.0mu}\scriptscriptstyle G}}
{\cal P}_s^{\mspace{1.0mu}\scriptscriptstyle G}
(m,-i {\partial}/ {\partial{q_{\scriptscriptstyle{\cal L}}}})\,
\int\limits_{-\infty}^{\infty}
(d^{\,n}p)^{\scriptscriptstyle{\cal N}}\,
i^{\lambda_{\scriptscriptstyle {\cal L}}}
\notag\\[-4pt]
\phantom{%
\cdot
\sum_{s=0}^{d^{\mspace{1.0mu}\scriptscriptstyle G}}
{\cal P}_s^{\mspace{1.0mu}\scriptscriptstyle G}
(m,-i
}%
\cdot\,
e^{%
-iM_{\epsilon}%
\,+\,
iW^{q_{\sscr {\cal L}}}_{{\sscr {\cal N}},{\sscr {\cal E}}}
}%
{\bigg|}_{%
\begin{subarray}{l}
q_l=0\\ \forall{l}\in{\cal L}
\end{subarray} },
\tag{$3.13$}\\ 
W^{q_{\sscr {\cal L}}}_{{\sscr {\cal N}},{\sscr {\cal E}}}\!:=
(p^{\sscr T}_{\sscr {\cal N}}\cdot
C_{\sscr\cal N\cal N}(\alpha)
p_{\sscr {\cal N}}) +
2(f^{\sscr T}_{\sscr {\cal N}}\cdot
p_{\sscr {\cal N}})
\notag\\[2pt]
\phantom{%
W^{q_{\sscr {\cal L}}}_{{\sscr {\cal N}},{\sscr {\cal E}}}\!\!:
}%
+
(k^{\sscr T}_{\sscr {\cal E}}\cdot
E_{\sscr\cal E\cal E}(j|\alpha)
k_{\sscr {\cal E}}) +
(q^{\sscr T}_{\sscr {\cal L}}\cdot
e_{\sscr\cal L\cal E}(j)
k_{\sscr {\cal E}}),
\notag\\[2pt]
f_{\sscr {\cal N}}\!:=
\Pi^{\sscr T}_{\sscr\cal E\cal N}(j|\alpha)
k_{\sscr {\cal E}} +
\tfrac{1}{2} e_{\sscr\cal L\cal N}^{\sscr T}
q_{\sscr {\cal L}},
\notag\\[2pt]
\delta^{\mspace{1.0mu}\sscr G}(k_{\sscr{\cal E}})\!:=
\delta\bigl(
{\textstyle\sum\limits}_{e\in{\cal E}}\,
e(v^{\ast})_ek_e
\bigr).
\notag
\end{gather}
The explicit forms and some properties of the matrices
$e_{\sscr\cal L\cal N}$,
$e_{\sscr\cal L\cal E}(j)$,
$C_{\sscr\cal N\cal N}(\alpha)$,
$E_{\sscr\cal E\cal E}(j|\alpha)$,
and
$\Pi_{\sscr\cal E\cal N}(j|\alpha)$
are given in Eqs.(5.1)-(5.4).

The change of variables $p_{\sscr {\cal N}}$ by means of the nondegenerate
linear transformation such that,
\begin{gather}
p_{\sscr {\cal N}} =
B_{\sscr\cal N\cal N}(\alpha){\tilde p}_{\sscr {\cal N}}-
B_{\sscr\cal N\cal N}(\alpha)B^{\,\sscr T}_{\sscr\cal N\cal N}(\alpha)
f_{\sscr {\cal N}},
\notag\\[2pt]
B^{\,\sscr T}_{\sscr\cal N\cal N}(\alpha)
C_{\sscr\cal N\cal N}(\alpha)
B_{\sscr\cal N\cal N}(\alpha)=1_{\sscr\cal N\cal N},
\notag\\[2pt]
B_{\sscr\cal N\cal N}(\alpha)
B^{\,\sscr T}_{\sscr\cal N\cal N}(\alpha)=
C^{-1}_{\sscr\cal N\cal N}(\alpha),
\notag\\[2pt]
\det B_{\sscr\cal N\cal N}(\alpha)=
[\det C_{\sscr\cal N\cal N}(\alpha)]^{-1/2}=:\!\Delta(\alpha)^{-1/2},
\notag\\[2pt]
(d^{\,n}p\,)^{\sscr {\cal N}}=
(d^{\,n}{\tilde p}\,)^{\sscr {\cal N}}
|\det B_{\sscr\cal N\cal N}(\alpha)|^{\,n}\,|\det g|^{\,\sscr |\cal N|}
\notag\\[2pt]
=
(d^{\,n}{\tilde p}\,)^{\sscr {\cal N}}/\Delta(\alpha)^{n/2},\
\det g=(-1)^q,
\tag{$3.14$} 
\end{gather}
to reduce Eqs.(3.13)-(3.14) to the form
\begin{gather}
I^{\scriptscriptstyle G}(m,k)^{\vec{r}\,}_{\epsilon}\!:=
\delta^{\mspace{1.0mu}\scriptscriptstyle G}(k_{\sscr{\cal E}})
c^{\mspace{1.0mu}\scriptscriptstyle G}
\int_{ {R}_{+}^{\scriptscriptstyle|{\cal L}|}(\vec{r}\,) }
\frac {dv^{\sscr G}(\alpha)} {\Delta^{n/2}}
\notag\\ 
\cdot
\sum_{s=0}^{d^{\mspace{1.0mu}\scriptscriptstyle G}}
{\cal P}_s^{\mspace{1.0mu}\scriptscriptstyle G}
(m,-i {\partial}/ {\partial{q_{\scriptscriptstyle{\cal L}}}})\,%
\int\limits_{-\infty}^{\infty}
(d^{\,n}{\tilde p}\,)^{\sscr {\cal N}}
e^{%
i({\tilde p}^{\mspace{1.0mu}\sscr T}_{\sscr {\cal N}}\mspace{1.0mu}\cdot
\mspace{1.0mu}{\tilde p}_{\sscr {\cal N}})
}%
\,i^{\lambda_{\scriptscriptstyle {\cal L}}}
\notag\\[-4pt]
\phantom{%
\cdot
\sum_{s=0}^{d^{\mspace{1.0mu}\scriptscriptstyle G}}
{\cal P}_s^{\mspace{1.0mu}\scriptscriptstyle G}
(m,-i
}%
\cdot\,
e^{%
-iM_{\epsilon}%
\,+\,
i{\widetilde W}^{q_{\sscr {\cal L}}}_{\sscr {\cal E}}
}%
{\bigg|}_{%
\begin{subarray}{l}
q_l=0\\ \forall{l}\in{\cal L}
\end{subarray} },
\tag{$3.15$}\\ 
{\widetilde W}^{q_{\sscr {\cal L}}}_{\sscr {\cal E}}\!:=
-
(f^{\sscr T}_{\sscr {\cal N}}\cdot
C^{-1}_{\sscr\cal N\cal N}(\alpha)
f_{\sscr {\cal N}}) +
\notag\\[2pt]
\phantom{%
{\widetilde W}^{q_{\sscr {\cal L}}}_{\sscr {\cal E}}\!\!:
}%
+
(k^{\sscr T}_{\sscr {\cal E}}\cdot
E_{\sscr\cal E\cal E}(j|\alpha)
k_{\sscr {\cal E}}) +
(q^{\sscr T}_{\sscr {\cal L}}\cdot
e_{\sscr\cal L\cal E}(j)
k_{\sscr {\cal E}})= 
\notag\\[2pt]
\phantom{%
{\widetilde W}^{q_{\sscr {\cal L}}}_{\sscr {\cal E}}\!\!:
}%
=
(k^{\sscr T}_{\sscr {\cal E}}\cdot
A_{\sscr\cal E\cal E}(j|\alpha)
k_{\sscr {\cal E}}) +
\notag\\[2pt]
\phantom{%
{\widetilde W}^{q_{\sscr {\cal L}}}_{\sscr {\cal E}}\!\!:
}%
+
(q^{\sscr T}_{\sscr {\cal L}}\cdot
Y_{\sscr\cal L\cal E}(j|\alpha)
k_{\sscr {\cal E}}) -
\tfrac{1}{4}
(q^{\sscr T}_{\sscr {\cal L}}\cdot
X_{\sscr\cal L\cal L}(\alpha)
q_{\sscr {\cal L}}).
\notag
\end{gather}

Taking into account the formula
\begin{gather*}
{\textstyle\int\limits}_{-\infty}^{\infty}
dt e^{\pm i t^2}%
=\pi^{1/2} e^{\pm i\pi/4},\
%
\end{gather*}
which is followed from~\cite{%
Bateman&Erdelyi1965},
Ch.\,1.5., eqs.(31-32), we find
\begin{gather}
{\textstyle\int\limits}_{-\infty}^{\infty}
d^{\,n}{\tilde p}_l
e^{i {\tilde p}_l^2}%
=\pi^{n/2} e^{i(p-q)\pi/4}=\pi^{n/2} i^{(p-n/2)},
\notag\\[2pt]
{\textstyle\int\limits}_{-\infty}^{\infty}
(d^{\,n}{\tilde p}\,)^{\sscr {\cal N}}
e^{%
i({\tilde p}^{\mspace{1.0mu}\sscr T}_{\sscr {\cal N}}\mspace{1.0mu}\cdot
\mspace{1.0mu}{\tilde p}_{\sscr {\cal N}})
}%
i^{\lambda_{\scriptscriptstyle {\cal L}}}
=
(\pi^{n/2}i^p)^{\sscr |{\cal N}|} i^{-\omega}.
\tag{$3.16$}
\end{gather}
So, all $n$-dimensional pseudo-Euclidean momenta integrations in
the $(p,q)$-metric are performed.
Thus, any FA (3.1) is led to the $\alpha$-parametric
representation in the fully-exponential form,
\begin{gather}
I^{\scriptscriptstyle G}(m,k)^{\vec{r}}_{\epsilon}\!:=
(2\pi)^n
\delta^{\mspace{1.0mu}\scriptscriptstyle G}(k_{\sscr{\cal E}})
a^{\mspace{1.0mu}\scriptscriptstyle G}
\int_{ {R}_{+}^{\scriptscriptstyle|{\cal L}|}(\vec{r}\,) }
\frac {dv^{\sscr G}(\alpha)} {\Delta^{n/2}}
\notag\\ 
\cdot
\sum_{s=0}^{d^{\mspace{1.0mu}\scriptscriptstyle G}}
{\cal P}_s^{\mspace{1.0mu}\scriptscriptstyle G}
(m,-i {\partial}/ {\partial{q_{\scriptscriptstyle{\cal L}}}})\,
e^{%
-iM_{\epsilon}%
\,+\,
i{\widetilde W}^{q_{\sscr {\cal L}}}_{\sscr {\cal E}}
}%
{\bigg|}_{%
\begin{subarray}{l}
q_l=0\\ \forall{l}\in{\cal L}
\end{subarray} },
\tag{$3.17$}\\[2pt]
{\widetilde W}^{q_{\sscr {\cal L}}}_{\sscr {\cal E}}\!:=
(k^{\sscr T}_{\sscr {\cal E}}\cdot
A_{\sscr\cal E\cal E}(j|\alpha)
k_{\sscr {\cal E}}) +
\notag\\[2pt]
\phantom{%
{\widetilde W}^{q_{\sscr {\cal L}}}_{\sscr {\cal E}}\!\!:
}%
+
(q^{\sscr T}_{\sscr {\cal L}}\cdot
Y_{\sscr\cal L\cal E}(j|\alpha)
k_{\sscr {\cal E}}) -
\tfrac{1}{4}
(q^{\sscr T}_{\sscr {\cal L}}\cdot
X_{\sscr\cal L\cal L}(\alpha)
q_{\sscr {\cal L}}),
\notag\\[2pt]
\,
a^{\mspace{1.0mu}\scriptscriptstyle G}\!:=
c^{\mspace{1.0mu}\scriptscriptstyle G}
(\pi^{n/2}i^p)^{\sscr |{\cal N}|} (2\pi)^{-n} i^{-\omega},\
|{\cal N}|=|{\cal C}|,
\notag
\end{gather}
where $n$-dimensional auxilary momenta, $q_l, l\in {\cal L}$,
still are used.
The explicit form and important properties of matrices
$A_{\sscr\cal E\cal E}(j|\alpha)$,
$Y_{\sscr\cal L\cal E}(j|\alpha)$,
and
$X_{\sscr\cal L\cal L}(\alpha)$
are given in Eqs.(5.5)-(5.13). Some properties of them are new.

Next, the following two operations must be carried out:
i) to differentiate the exponential function
$\exp\{%
i(q^{\sscr T}_{\sscr {\cal L}}\cdot
Y_{\sscr\cal L\cal E}(j|\alpha)
k_{\sscr {\cal E}}) -
%
{i}/{4}\,
(q^{\sscr T}_{\sscr {\cal L}}\cdot
X_{\sscr\cal L\cal L}(\alpha)
q_{\sscr {\cal L}})
\}$
by means of the $s$-homogeneous differential polynomials
${\cal P}_s^{\mspace{1.0mu}\scriptscriptstyle G}
(m,-i {\partial}/ {\partial{q_{\scriptscriptstyle{\cal L}}}})$
in $-i {\partial}/ {\partial{q_{l\sigma}} }$,
$l\in {\cal L}$, $\sigma\in \{1, \ldots, n\}$,
$1\leq s \leq d^{\mspace{1.0mu}\scriptscriptstyle G}$;
ii) to put
${q_{l\sigma}}=0, \forall l\in {\cal L}$,
$\forall \sigma\in \{1, \ldots, n\}$,
and
$\forall s\in \{0,1, \ldots, d^{\mspace{1.0mu}\scriptscriptstyle G}\}$.
Finally, we obtain the important $\alpha$-parametric representation
for the general FA (3.1),
\begin{gather}
I^{\scriptscriptstyle G}(m,k)^{\vec{r}}_{\epsilon}\!:=
(2\pi)^n
\delta^{\mspace{1.0mu}\scriptscriptstyle G}(k_{\sscr{\cal E}})
b^{\mspace{1.0mu}\scriptscriptstyle G}
\int_{ {R}_{+}^{\scriptscriptstyle|{\cal L}|}(\vec{r}\,) }
{dv^{\sscr G}(\alpha)}
\notag\\ 
\cdot
\frac {1}{\Delta^{n/2}}
\sum_{s=0}^{d^{\mspace{1.0mu}\scriptscriptstyle G}}
\sum_{j=0}^{[s/2]}
{\cal P}^{\scriptscriptstyle G}_{sj}(m,\alpha,k)
i^{-\omega-j}
e^{-iM_{\epsilon}+iA},
\tag{$3.18$}\\[2pt]
A\equiv A(\alpha,k)\!:=
(k^{\sscr T}_{\sscr {\cal E}}\cdot
A_{\sscr\cal E\cal E}(j|\alpha)
k_{\sscr {\cal E}}),
\notag\\[2pt]
\,
b^{\mspace{1.0mu}\scriptscriptstyle G}\!:=
c^{\mspace{1.0mu}\scriptscriptstyle G}
(\pi^{n/2}i^p)^{\sscr |{\cal C}|} (2\pi)^{-n},\
|{\cal C}|=|{\cal N}|,
\notag\\[2pt]
{\cal P}^{\scriptscriptstyle G}_{sj}(m,\rho\alpha,\tau k)=
\rho^{-j}\tau^{s-2j}\,
{\cal P}^{\scriptscriptstyle G}_{sj}(m,\alpha,k).
\notag
\end{gather}
Here $[s/2]$ means the largest integer $\leq s/2$,
i.e., the integer part of the $s/2$;
the quadratic Kirhchhoff form $A(\alpha,k)$ in external momenta
$k_e$, $e\in {\cal E}$, and the Kirhchhoff determinant
$\Delta(\alpha)\!:=\det C_{\sscr\cal N\cal N}(\alpha)$
are defined by the topological structure of a graph $G$
and are homogeneous functions of the first and $|{\cal C}|$th
degrees in $\alpha$, respectively, see Sec.5 more detail.
The quantities
${\cal P}^{\scriptscriptstyle G}_{sj}(m,\alpha,k)$
are homogeneous $k$-polynomials in external momenta
$k_e$, $e\in {\cal E}$, of the degree $s-2j$, $j=0,1,\ldots,[s/2]$.
They are $\alpha$-parametric images of homogeneous polynomials
${\cal P}^{\scriptscriptstyle G}_{s}(m,p)$.
Each monomial of
${\cal P}^{\scriptscriptstyle G}_{sj}(m,\alpha,k)$
is a product of $s-2j$ linear Kirchhoff forms
$Y_l(\alpha,k)\!:=\sum_{e\in {\cal E}}Y_{le}(\alpha)k_e$
and $j$ line-correlator functions
$X_{ll'}(\alpha),\,l,l'\in {\cal L}$, of a graph $G$.
Parametric functions
$Y_l(\alpha,k)$
and
$X_{ll'}(\alpha)$
are homogeneous functions of the 0th and (-1)st degree
in $\alpha$, respectively.
See Sec.4 and 5 more detail.

By introducing the new variables,
\begin{gather}
\alpha_l=\rho{\alpha}_l',\ \ \forall l\in{\cal L}/j,\ \
\alpha_j=
\rho\bigl(1-{\textstyle\sum\limits}_{l\in{\cal L}/j}{\alpha}_l'\bigr),
\notag\\[2pt]
\sum_{l\in{\cal L}}{\alpha}_l'=1,\ \
\prod_{l\in{\cal L}}{d\alpha}_l=
\rho^{{\sscr |{\cal L}|}-1} d\rho
\prod_{l\in{\cal L}/j}{d\alpha}_l',
\notag\\[-1pt]
{R}_{+}^{\scriptscriptstyle|{\cal L}|}(\vec{r}\,)
\rightarrow
{R}_{+}^{1}(r) \times \Sigma^{|{\cal L}|-1},\ r>0,
\tag{$3.19$} 
\end{gather}
and assuming that $r_l=r>0,\, \forall l\in {\cal L}$,
we can perform the integration over the variable $\rho$,
$0< r\leq \rho \leq\infty$,
by using~\cite{%
Bateman&Erdelyi1969},
see Ch.\,6.3., eq.(3) or Ch.\,6.5., eq.(29),
\begin{gather}
I^{\scriptscriptstyle G}(m,k)^{r}_{\epsilon}\!:=
(2\pi)^n
\delta^{\mspace{1.0mu}\scriptscriptstyle G}(k_{\sscr{\cal E}})
b^{\mspace{1.0mu}\scriptscriptstyle G}
\int\limits_{\Sigma^{|{\cal L}|-1}}
\frac {d\mu^{\scriptscriptstyle G}(\alpha)} {\Delta^{n/2}}
\notag\\[-1pt]
\cdot
\sum_{s=0}^{d^{\mspace{1.0mu}\scriptscriptstyle G}}
\sum_{j=0}^{[s/2]}
{\cal P}^{\scriptscriptstyle G}_{sj}(m,\alpha,k)
{\cal F}_{sj}^{\,r}(\omega;V_{\epsilon}),
\notag\\[-1pt]
{\cal F}_{sj}^{\mspace{1.0mu}r}(\omega;V_{\epsilon})\!:=
i^{-\omega-j}
\int\limits_r^\infty
d\rho\,\rho^{-\omega-j-1}
e^{%
-i\rho\,V_{\epsilon}
}%
\notag\\[-1pt]
=
V_{\epsilon}^{\,\omega+j}\Gamma(-\omega-j;irV_{\epsilon}),\ \
V_{\epsilon}\!:= M_{\epsilon}-A,
\tag{$3.20$}\\[-1pt]
\omega\!:=
({n}/{2})|{\cal C}|
- \lambda_{\sscr {\cal L}},\ \
\,
b^{\mspace{1.0mu}\sscr G}\!:=
c^{\mspace{1.0mu}\sscr G}
(\pi^{n/2}i^p)^{\sscr |{\cal C}|} (2\pi)^{-n}.
\notag
\end{gather}
Here $p$ involved in the
$b^{\scriptscriptstyle G}$
is the number of positive squares in
a space-time metric $g^{\mu\nu}$.
The integration measure
$d\mu^{\scriptscriptstyle G}(\alpha)$
and the integration domain
$\Sigma^{|{\cal L}|-1}$
of the simplex type are defined as
\begin{gather}
d\mu^{\scriptscriptstyle G}(\alpha)\!:=
\delta\bigl(1-\sum_{l\in {\cal L}}\alpha_l\bigr)
\prod_{l\in {\cal L}}
\biggl(
\frac {d\alpha_l\,\alpha_l^{\lambda_l-1}} {\Gamma(\lambda_l)}
\biggr),
\notag\\[-1pt]
\Sigma^{|{\cal L}|-1}\!:=
\{
\alpha_l|\,\alpha_l\ge\,0,\
\forall l\in {\cal L},\ \sum_{l\in {\cal L}}\alpha_l=1
\}.
\tag{$3.21$}
\end{gather}
In Eq.(3.21) $\Gamma(\alpha;x)$ is one in a two
incomplete gamma functions appearing in the decomposition,
$\Gamma(\alpha)=\Gamma(\alpha;x) + \gamma(\alpha;x)$,
see~\cite{%
Bateman&Erdelyi1974},
Ch.\,9.1., eqs.(1-2),
such that at Re\,$\alpha>0$, $\Gamma(\alpha;0)=\Gamma(\alpha)$,
$\gamma(\alpha;0)=0$,
where $\Gamma(\alpha)$ is an ordinary gamma function.

It is useful to remark that
we actually have the regularization which combines three ones:
i) the Hepp regularization~\cite{%
Hepp1966},
(due to the change in the region of integration over
auxilary variable $\rho$);
ii) the analytic regularization~\cite{%
Speer1969},
(due to the complexification of the parameter
$\lambda_{\sscr {\cal L}}$,
half-degree of the denominator polynomial);
iii) the dimensional regularization~\cite{%
Kucheryavy1992,
Kucheryavy1993},
(due to the complexification of the parameter $n$,
the space-time dimension).
Recall, that
$\lambda_{\sscr {\cal L}}$ and $n$
are constituents of $\omega$.

For convergent FAs the quantities
${\omega+j}<0$,
$\forall j\in \{0,1,\ldots,[d^{\,\sscr G}/2\}$
and there exists the limit $r\to 0$.
After passing to the limit $r\to 0$ in Eq.(3.20) we obtain,
\begin{gather}
I^{\scriptscriptstyle G}(m,k)_{\epsilon}\!:=
(2\pi)^n
\delta^{\mspace{1.0mu}\scriptscriptstyle G}(k_{\sscr{\cal E}})
b^{\mspace{1.0mu}\scriptscriptstyle G}
\int\limits_{\Sigma^{|{\cal L}|-1}}
\frac {d\mu^{\scriptscriptstyle G}(\alpha)} {\Delta^{n/2}}
\notag\\[-1pt]
\cdot
\sum_{s=0}^{d^{\mspace{1.0mu}\scriptscriptstyle G}}
\sum_{j=0}^{[s/2]}
{\cal P}^{\scriptscriptstyle G}_{sj}(m,\alpha,k)
{\cal F}_{sj}(\omega;M_{\epsilon},A),
\notag\\[-1pt]
{\cal F}_{sj}(\omega;M_{\epsilon},A)\!:=
i^{-\omega-j}
\int\limits_0^\infty
d\rho\,\rho^{-\omega-j-1}
e^{%
-i\rho\,(M_{\epsilon} - A)
}%
\notag\\[-1pt]
=
M_{\epsilon}^{\,\omega+j}(1-Z_{\epsilon})^{\,\omega+j}
\Gamma(-\omega-j)
\tag{$3.22$}\\[-1pt]
=
M_{\epsilon}^{\,\omega+j}
\sum_{k=0}^{\infty}
\Gamma(-\,\omega-j+k)\,
\dfrac
{Z_{\epsilon}^k} {k!}, \ \ \:
Z_{\epsilon}\!:=A/M_{\epsilon}.
\notag 
\end{gather}
It is easily verified that basic functions
${\cal F}_{sj}(\omega;M_{\epsilon},A)$
satisfy Eqs.(2.1), (2.3), (2.11), and Eqs.(2.6)-(2.8).

In the case of divergent FAs for which
${\omega+j}\geq0$
at least for one
$j\in \{0,1,\ldots,[d^{\,\sscr G}/2\}$,
the limit $r\to 0$ does not exist.
In this case, the expressions (3.6r)-(3.6s) strictly defined
in the region
${R}_{+}^{\scriptscriptstyle|{\cal L}|}(r)\!:=
{ {R}_{+}^1(r) }\times\Sigma^{|{\cal L}|-1}$
must be made meaningful in a wider region
${R}_{+}^{\scriptscriptstyle|{\cal L}|}\!:=
{R}_{+}^1 \times\Sigma^{|{\cal L}|-1}$,
where ${R}_{+}^1\!:={R}_{+}^1(r)|_{r=0}$.

The Bogoliubov-Parasiuk subtraction procedure applied for
this purpose replaces
$I^{\scriptscriptstyle G}(m,k)_{\epsilon}^r$
by
\begin{gather}
(R_{\mspace{1.0mu}\scriptscriptstyle 0}^{\mspace{1.0mu}\nu}
I)^{\scriptscriptstyle G}(m,k)_{\epsilon}
=
(2\pi)^n
\delta^{\mspace{1.0mu}\scriptscriptstyle G}(k_{\sscr{\cal E}})
b^{\mspace{1.0mu}\scriptscriptstyle G}
\notag\\[-1pt]
\phantom{%
(R_{\mspace{1.0mu}\scriptscriptstyle 0}^{\mspace{1.0mu}\nu}
}%
\cdot
\int_{ {R}_{+}^{\scriptscriptstyle|{\cal L}|} }
{dv^{\sscr G}(\alpha)}
(R_{\mspace{1.0mu}\scriptscriptstyle 0}^{\mspace{1.0mu}\nu}
{\cal I})^{\scriptscriptstyle G}(m,\alpha,k)_{\epsilon},
\tag{$3.23$}\\[-1pt]
(R_{\mspace{1.0mu}\scriptscriptstyle 0}^{\mspace{1.0mu}\nu}
{\cal I})^{\scriptscriptstyle G}(m,\alpha,k)_{\epsilon}\!:=
\notag\\[-1pt]
\!:=
{\cal I}^{\scriptscriptstyle G}(m,\alpha,k)_{\epsilon}
-
\sum_{\beta=0}^{\nu}
\frac {1}{\beta\,!}
\frac {\partial^{\,\beta}}{\partial\tau^{\beta}}
{\cal I}^{\scriptscriptstyle G}(m,\alpha,\tau k)_{\epsilon}
\biggl\arrowvert_{\tau=0},
\notag\\[-1pt]                   
=
\frac {1}{\nu\,!}
\int_0^1 d\tau (1-\tau)^{\nu}
\frac {\partial^{\,\nu+1}}{\partial\tau^{\nu+1}}
{\cal I}^{\scriptscriptstyle G}(m,\alpha,\tau k)_{\epsilon}.
\tag{$3.24$}                     
\end{gather}
where subtraction operations under the integral sign are performed
by using the Schl{\"o}milch integro-differential formula,
see the 2nd line of Eq.(3.24),
for the remainder term of Maclaurin's series.
Firstly, this formula was applied explicitly to the FAs in the
Parasiuk paper~\cite{%
Parasiuk1960}.
Although this expression guarantees a compact representation of the
subtraction procedure, it is, nevertheless, inconvenient for computational
purposes, because it involves additional integration and differentiations
in the integrand.
The expression in the 1st line of Eq.(3.24) is all the more inconvenient
for these purposes,
since every term on the right-hand side of one may be assosiated with a
divergent integral.

At the same time the algorithm proposed and applied in~\cite{%
Kucheryavy1974,
Kucheryavy1977,
Kucheryavy1979,
Kucheryavy1982,
Kucheryavy1982a,
Kucheryavy1983,
Kucheryavy1983a,
Kucheryavy1987,
Kucheryavy1991}
is based on the observation,
see~\cite{%
Bateman&Erdelyi1974},
Ch.\,9.2., eqs.(16,\,17,\,18), that,
\begin{gather}
e^x - \sum_{k=0}^{\nu_{sj}} \frac{x^k}{k\,!}=
e^x
\tilde{\gamma}(1+\nu_{sj};x),\
\tilde{\gamma}(\alpha;x)\!:=\frac {\gamma(\alpha;x)} {\Gamma(\alpha)},
\notag\\[1pt]
\sum_{k=0}^{\nu_{sj}}\frac {x^k} {k\,!}=
e^x
\tilde{\Gamma}(1+\nu_{sj};x),\
\tilde{\Gamma}(\alpha;x)\!:=\frac {\Gamma(\alpha;x)} {\Gamma(\alpha)},
\notag\\[1pt]
\tilde{\gamma}(\alpha;x) + \tilde{\Gamma}(\alpha;x)=1.
\tag{$3.25$}
\end{gather}
Now, if we use:
the explicit form of the integrand in Eq.(3.18);
the homogeneous properties for parametric functions in
$k_e, e\in {\cal E}$, see Eqs.(5.8);
the 1st line of Eq.(3.24);
the 1st line of Eq.(3.25);
and the following useful relation,
\begin{gather}
\sum_{\beta=0}^{\nu}
\frac {1} {\beta\,!}
\frac {\partial^{\,\beta}}{\partial\tau^{\beta}}
\bigl\{
\tau^{s-2j}
e^{i\tau^2 A}
\bigr\}
\bigl\arrowvert_{\tau=0}=
\sum_{k=0}^{\nu_{sj}} \frac{(i A)^k}{k\,!},
\notag\\[1pt]
\nu_{sj}\!:=[(\nu-s)/2] +j,
\tag{$3.26$}
\end{gather}
we arrive at the multiplicative realization of the subtraction procedure
in the integrand of Eq.(3.23) for the regular value of general FA (3.1),
\begin{gather}
(R_{\mspace{1.0mu}\scriptscriptstyle 0}^{\mspace{1.0mu}\nu}
{\cal I})^{\scriptscriptstyle G}(m,\alpha,k)_{\epsilon}=
\frac {1}{\Delta^{n/2}}
\sum_{s=0}^{d^{\mspace{1.0mu}\scriptscriptstyle G}}
\sum_{j=0}^{[s/2]}
\notag\\ 
\cdot
{\cal P}^{\scriptscriptstyle G}_{sj}(m,\alpha,k)
i^{-\omega-j}
e^{-iV_{\epsilon}}\tilde{\gamma}(1+\nu_{sj};iA).
\tag{$3.27$}
\end{gather}
The integral in Eq.(3.23) with the integrand (3.27) at
$\nu\geq\nu^{\sscr G}$ is now well-defined in the domain
${R}_{+}^{\sscr |{\cal L}|}$.
The substitution of (3.27) into the integral (3.23) and the
change of variables in integration according to Eq.(3.19)
give rise to the expression
\begin{gather}
(R_{\mspace{1.0mu}\scriptscriptstyle 0}^{\mspace{1.0mu}\nu}
I)^{\scriptscriptstyle G}(m,k)_{\epsilon}= 
(2\pi)^n\delta^{\scriptscriptstyle G}(k)\,b^{\scriptscriptstyle G}
\int\limits_{\Sigma^{|{\cal L}|-1}}
\frac {d\mu^{\scriptscriptstyle G}(\alpha)} {\Delta^{n/2}}
\notag\\ 
\cdot
\sum_{s=0}^{d^{\mspace{1.0mu}\scriptscriptstyle G}}
\sum_{j=0}^{[s/2]}
{\cal P}^{\scriptscriptstyle G}_{sj}(m,\alpha,k)
(R_{\mspace{1.0mu}\scriptscriptstyle 0}^{\mspace{1.0mu}\nu}
{\cal F})_{sj}(\omega;M_{\epsilon},A),
\notag\\ 
(R_{\mspace{1.0mu}\scriptscriptstyle 0}^{\mspace{1.0mu}\nu}
{\cal F})_{sj}(\omega;M_{\epsilon},A)\!:=  
i^{-\omega-j}
\tag{$3.28$}\\
\cdot
\int_0^\infty
d\rho\,\rho^{-\omega-j-1}
e^{
-i\rho(M_{\epsilon} - A)
}\,%
\tilde{\gamma}(1+\nu_{sj};i\rho\,A)=
\notag\\ 
=
M_{\epsilon}^{\,\omega+j}                                 q
\frac {\Gamma(\lambda_{sj})} {\Gamma(2+\nu_{sj})}
Z_{\epsilon}^{\,1+\nu_{sj}}
{_2\mspace{1.0mu}\!F_1}(1,\lambda_{sj};\,2+\nu_{sj};\,Z_{\epsilon}), 
\notag\\ 
\nu_{sj}\!:=[(\nu-s)/2] +j,\
\lambda_{sj}\!:=-\,\omega-j+1+\nu_{sj}.
\notag 
\end{gather}
The integration over $\rho$ in (3.28) is performed
with the use of the formula, see~\cite{%
Bateman&Erdelyi1970},
Ch.\,17.3., eq.(15),
\begin{gather}
\int_0^\infty
dx\,
x^{\mu-1}
e^{-vx}
\tilde{\gamma}(\nu;a\,x)=
\notag\\ 
=
\frac {a^\nu \Gamma(\mu+\nu)} {(a+v)^{\mu+\nu}\Gamma(1+\nu)}\,
{_2\mspace{1.0mu}\!F_1}\bigl(1,\mu+\nu;\,1+\nu;\,\frac{a}{a+v}\bigr),
\notag\\ 
{\rm Re}\,(a+v)>0,\ {\rm Re}\,v>0,\ {\rm Re}\,(\mu+\nu)>0.
\notag
\end{gather}

\smallskip
{\bf 3.4}
So, using properties of special functions substantially,
the author has obtained~\cite{
Kucheryavy1974,
Kucheryavy1977,
Kucheryavy1982,
Kucheryavy1983,
Kucheryavy1983a,
Kucheryavy1991,
Kucheryavy1991a,
Kucheryavy1991b,
Kucheryavy2002,
Kucheryavy2004}
high-efficient formulas which realize an analytical continuation
(in the variables
$\omega^{\mspace{1.0mu}\scriptscriptstyle G}$
and
$\nu^{\mspace{1.0mu}\scriptscriptstyle G}$)
of the FAs which are represented first in Eqs.(3.1)-(3.3)
by UV-divergent integrals,
and are given finally in Eqs.(3.28)-(3.30) as convergent ones.
As a result, we have the following
$\alpha$-parametric integral representation,
\begin{align}
&
\begin{bmatrix}
I^{\scriptscriptstyle G}(m,k)_{\epsilon}\\[2pt]
(R_{\mspace{1.0mu}\scriptscriptstyle 0}^{\mspace{1.0mu}\nu}
I)^{\scriptscriptstyle G}(m,k)_{\epsilon}
\end{bmatrix} =
(2\pi)^n\delta^{\scriptscriptstyle G}(k)\,b^{\scriptscriptstyle G}
\int\limits_{\Sigma^{|{\cal L}|-1}}
\frac {d\mu^{\scriptscriptstyle G}(\alpha)} {\Delta^{n/2}}
\notag\\ 
&
\cdot
\sum_{s=0}^{d^{\mspace{1.0mu}\scriptscriptstyle G}}
\sum_{j=0}^{[s/2]}
{\cal P}^{\scriptscriptstyle G}_{sj}(m,\alpha,k)
\!
\begin{bmatrix}
{\cal F}_{sj}(\omega;M_{\epsilon},A)\\[2pt]
(R_{\mspace{1.0mu}\scriptscriptstyle 0}^{\mspace{1.0mu}\nu}
{\cal F})_{sj}(\omega;M_{\epsilon},A)
\end{bmatrix}
\!,
\tag{$3.29$}
\end{align}
for convergent or dimensionally regularized value
$I^{\scriptscriptstyle G}(m,k)_{\epsilon}$,
and for regular value
$(R_{\mspace{1.0mu}\scriptscriptstyle 0}^{\mspace{1.0mu}\nu}
I)^{\scriptscriptstyle G}(m,k)_{\epsilon}$
of the integral (3.1).
The subscript $0$ and superscript $\nu$ on $R$ indicate that
$(R_{\mspace{1.0mu}\scriptscriptstyle 0}^{\mspace{1.0mu}\nu}
I)^{\scriptscriptstyle G}(m,k)_{\epsilon}$
is the regular function in the vicinity of zero values of
external momenta
$k_e$, $e\in{\cal E}$,
and is evaluated for an renormalization index
$\nu=\nu^{\scriptscriptstyle G}$.

The explicit forms of basic functions
${\cal F}_{sj}(\omega;M_{\epsilon},A)$ and
$(R_{\mspace{1.0mu}\scriptscriptstyle 0}^{\mspace{1.0mu}\nu}
{\cal F})_{sj}(\omega;M_{\epsilon},A)$
are as follows:
\begin{gather}
{\cal F}_{sj}(\omega;M_{\epsilon},A)\!:=
M_{\epsilon}^{\,\omega+j}
(1-Z_{\epsilon})^{\,\omega+j}\,\Gamma(-\,\omega-j)
\notag\\[1pt]
=
M_{\epsilon}^{\,\omega+j}
\sum_{k=0}^{\infty}
\Gamma(-\,\omega-j+k)\,
\dfrac
{Z_{\epsilon}^k} {k!}, \ \ \:
Z_{\epsilon}\!:=A/M_{\epsilon},
\notag\\[0pt]
(R_{\mspace{1.0mu}\scriptscriptstyle 0}^{\mspace{1.0mu}\nu}
{\cal F})_{sj}(\omega;M_{\epsilon},A)\!:=
M_{\epsilon}^{\,\omega+j}\,
{\Gamma(\lambda_{sj})}/ {\Gamma(2+\nu_{sj})}\cdot
\notag\\[4pt]
\cdot
Z_{\epsilon}^{\,1+\nu_{sj}}
{_2\mspace{1.0mu}\!F_1}(1,\lambda_{sj};\,2+\nu_{sj};\,Z_{\epsilon})
\notag\\[1pt]
=
M_{\epsilon}^{\,\omega+j}
\sum_{k=1+\nu_{sj}}^{\infty}
\Gamma(-\,\omega-j+k)\,
\dfrac
{Z_{\epsilon}^k} {k!}, \ \:
\tag{$3.30$}\\[0pt]
\nu_{sj}\!:=
[(\nu-s)/2]+j=
[\omega]+j+\sigma_s,
\notag\\[2pt]
\lambda_{sj}\!:=
-\,\omega-j+1+\nu_{sj}=
1-\delta_n\delta_{\scriptscriptstyle |{\cal C}|}/2+\sigma_s,
\notag\\[2pt]
[\omega]\!:=
r_n|{\cal C}|+
\delta_n r_{\scriptscriptstyle |{\cal C}|}-
\lambda,\ \
\omega=[\omega]+
\delta_n \delta_{\scriptscriptstyle |{\cal C}|}/2,
\notag\\[2pt]
\sigma_s\!:=
[(\delta_n\delta_{\scriptscriptstyle |{\cal C}|}+d-s)/2],\ \
|{\cal C}|=
2r_{\scriptscriptstyle |{\cal C}|}+
\delta_{\scriptscriptstyle |{\cal C}|},
\notag\\[2pt]
\nu = \nu^{\scriptscriptstyle G},\
\omega = \omega^{\scriptscriptstyle G},\
\lambda = \lambda^{\scriptscriptstyle G},\
d=d^{\scriptscriptstyle G}.
\notag
\end{gather}
The $[(\nu-s)/2]$, $[(\nu+1-s)/2]$, and $[\omega]$
in Eqs.(3.28)-(3.31) are the integer parts of the
$(\nu-s)/2$, $(\nu+1-s)/2$, and $\omega$, respectively.
The subscripts $(s,j)$ on
${\cal F}_{sj}$
and
$(R_{\mspace{1.0mu}\scriptscriptstyle 0}^{\mspace{1.0mu}\nu}
{\cal F})_{sj}$
just mean that these functions are attached
to the homogeneous $k$-polynomials
${\cal P}^{\scriptscriptstyle G}_{sj}(\alpha,m,k)$
of the degree $s-2j$, $j=0,\ldots,[s/2]$, in external momenta
$k_e$, $e\in{\cal E}$.
The latter are $\alpha$-images of the homogeneous $p$-polynomials
${\cal P}^{\scriptscriptstyle G}_{s}(m,p)$
of the degree $s$ appearing in
${\cal P}^{\scriptscriptstyle G}(m,p)$, see Eqs.(3.2).
The $k$-polynomials
${\cal P}^{\scriptscriptstyle G}_{sj}(\alpha,m,k)$
are constructed by means of $\alpha$-parametric functions
$Y_l(\alpha,k)$ and $X_{ll'}(\alpha)$, $l,l'\in {\cal L}$.
The efficient and universal algorithm of building
${\cal P}^{\scriptscriptstyle G}_{sj}(\alpha,m,k)$
is presented in Sec.4.
The $\alpha$-parametric functions
$M_{\epsilon}\equiv M(m,\alpha)_{\epsilon}$
and
$A\equiv A(\alpha,k)$,
incoming in Eqs.(3.30) are defined in Eqs.(3.9) and (3.18), respectively.
The $M(m,\alpha)_{\epsilon}$ is the linear form in
the square of internal masses with $i\epsilon$-damping.
The functions
$A(\alpha,k)$ and $Y_l(\alpha,k)$
are known as the quadratic and linear Kirchhoff forms
in external momenta,
$k_e$, $e\in{\cal E}$.
The function $\Delta\equiv\Delta(\alpha)$
is the Kirchhoff determinant, and
the $X_{ll'}(\alpha)$ are the line-correlator functions.
The high-efficient and universal algorithm of finding
$\alpha$-parametric functions
$A(\alpha,k)$, $Y_l(\alpha,k)$, $X_{ll'}(\alpha)$,
and $\Delta(\alpha)$
is given in Sec.5.

\smallskip
{\bf 3.5}
Investigation of complicated tangle of problems
associated on the one hand with renormali\-zation methods and
on the other with conserved and broken symmetries,
the Ward identities behavior,
the Schwinger terms contributions,
and quantum ano\-malies
requires of finding renormalized FAs for different divergence indices.
For example, amplitudes involved in the Ward identities
have divergence indices
$\nu^{\scriptscriptstyle G}$ and
$\nu^{\scriptscriptstyle G}+1$.

Regular values
$(R_{\mspace{1.0mu}\scriptscriptstyle 0}^{\mspace{1.0mu}\nu+1}\!
I)^{\scriptscriptstyle G}(m,k)_{\epsilon}$
calculated for the renormalization index
$\nu^{\scriptscriptstyle G}+1$,
once again, have form of Eq.(3.29) but with another basic functions
$(R_{\mspace{1.0mu}\scriptscriptstyle 0}^{\mspace{1.0mu}\nu+1}\!
{\cal F})_{sj}$:
\begin{gather}
(R_{\mspace{1.0mu}\scriptscriptstyle 0}^{\mspace{1.0mu}\nu+1}\!
{\cal F})_{sj}\!:=
M_{\epsilon}^{\,\omega+j}\,
{\Gamma(\lambda_{sj}^1)}/ {\Gamma(2+\nu_{sj}^1)}
\notag\\[2pt]
\phantom{%
(R_{\mspace{1.0mu}\scriptscriptstyle 0}^{\mspace{1.0mu}\nu+1}\!
{\cal F})
}%
\cdot
Z_{\epsilon}^{\,1+\nu_{sj}^1}
{_2\mspace{1.0mu}\!F_1}(1,\lambda_{sj}^1;\,2+\nu_{sj}^1;\,Z_{\epsilon}),
\tag{$3.31$}\\[2pt]
\nu_{sj}^1\!:=
[(\nu+1-s)/2]+j=
[\omega]+\sigma_s^1+j,
\notag\\[2pt]
\lambda_{sj}^1\!:=
-\,\omega-j+1+\nu_{sj}^1=
1+\sigma_s^1-\delta_n\delta_{\scriptscriptstyle |{\cal C}|}/2,
\notag\\[2pt]
\sigma_s^1\!:=
[(\delta_n\delta_{\scriptscriptstyle |{\cal C}|}+d+1-s)/2].
\notag
\end{gather}
In general,
$(R_{\mspace{1.0mu}\scriptscriptstyle 0}^{\mspace{1.0mu}\nu+1}\!
{\cal F})_{sj}\ne
(R_{\mspace{1.0mu}\scriptscriptstyle 0}^{\mspace{1.0mu}\nu}
{\cal F})_{sj}$,
as far as $\nu_{sj}^1\ne \nu_{sj}$. Difference between them is the
important quantity
\begin{gather}
(\Delta_{\mspace{1.0mu}\scriptscriptstyle 0}
^{\mspace{1.0mu}(\nu+1,\nu)}\!
{\cal F})_{sj}\!:=
(R_{\mspace{1.0mu}\scriptscriptstyle 0}^{\mspace{1.0mu}\nu+1}\!
{\cal F})_{sj}-
(R_{\mspace{1.0mu}\scriptscriptstyle 0}^{\mspace{1.0mu}\nu}
{\cal F})_{sj}
\notag\\[3pt]
=
-\,\Theta_{sj}^{\mspace{1.0mu}(\nu+1,\nu)}\,
\dfrac
{\Gamma(\lambda_{sj})} {\Gamma(2+\nu_{sj})}\,
M_{\epsilon}^{\,\omega+j}\,Z_{\epsilon}^{\,1+\nu_{sj}},
\tag{$3.32$}\\[3pt]
\Theta_{sj}^{\mspace{1.0mu}(\nu+1,\nu)}\!:=
H_{\scriptscriptstyle +}(\nu_{sj}^1)
\theta_s^{\mspace{1.0mu}(\nu+1,\nu)},
\notag\\[3pt]
\theta_s^{\mspace{1.0mu}(\nu+1,\nu)}\!:=
\nu_{sj}^1-\nu_{sj}=
\sigma_s^1-\sigma_s=
\mid{\delta_{\nu}-\delta_s}\mid,
\notag\\[3pt]
\nu=2r_{\nu} + \delta_{\nu}, \quad
s=2r_s + \delta_s, \quad
\nu,s\in \{0\cup\mathbb{N}_{\sscr{+}}\},
\notag
\end{gather}
where $H_{\scriptscriptstyle +}(x)$
is the Heaviside step function such that
$H_{\scriptscriptstyle +}(x)=0,\ x<0$,\
$H_{\scriptscriptstyle +}(x)=1,\ x\ge\,0$,
and
$\delta_\nu, \delta_s\!:=\nu({\rm mod}\,2), s({\rm mod}\,2)=0,1$.
It is this quantity that permits to obtain some efficient formulas
for calculating of the {\it quantum corrections} (QCs)
(i.e., quantum anomalies) to the canonical Ward identities (CWIs)
of the most general kind. For example, to the Ward identities
involving canonically non-conserved vector and (or) axial-vector
currents for nondegenerate fermion systems (i.e., for systems with
different fermion masses).
Another a very useful quantity that is produced
by differences
\begin{gather}
(\Delta_{\mspace{1.0mu}\scriptscriptstyle 0}
^{\mspace{1.0mu}(\nu+2,\nu)}\!
{\cal F})_{sj}\!:=
(R_{\mspace{1.0mu}\scriptscriptstyle 0}^{\mspace{1.0mu}\nu+2}\!
{\cal F})_{sj}-
(R_{\mspace{1.0mu}\scriptscriptstyle 0}^{\mspace{1.0mu}\nu}
{\cal F})_{sj}
\notag\\[3pt]
=
(R_{\mspace{1.0mu}\scriptscriptstyle 0}^{\mspace{1.0mu}\nu}\!
{\cal F})_{s-2,j}-
(R_{\mspace{1.0mu}\scriptscriptstyle 0}^{\mspace{1.0mu}\nu-2}
{\cal F})_{s-2,j}=
\notag\\[3pt]
=
(R_{\mspace{1.0mu}\scriptscriptstyle 0}^{\mspace{1.0mu}\nu}
{\cal F})_{s-2,j}-
(R_{\mspace{1.0mu}\scriptscriptstyle 0}^{\mspace{1.0mu}\nu}
{\cal F})_{sj}=
\notag\\[3pt]
=
-\,H_{\scriptscriptstyle +}(1+\nu_{sj})
\dfrac
{\Gamma(\lambda_{sj})} {\Gamma(2+\nu_{sj})}\,
M_{\epsilon}^{\,\omega+j}\,Z_{\epsilon}^{\,1+\nu_{sj}},
\tag{$3.33$}
\end{gather}
is closely related with
$(\Delta_{\mspace{1.0mu}\scriptscriptstyle 0}^{\mspace{1.0mu}(\nu+1,\nu)}\!
{\cal F})_{sj}$.

\smallskip
{\bf 3.6}
The expressions given by Eqs.(3.28)-(3.30) have two the
very important properties.

Firstly, they describe both divergent and convergent FAs
in the unified manner.
Really, due to properties~\cite{%
Bateman&Erdelyi1965}\
Ch.\;2.8, eqs.(4,\,19), i.e.,
${_2\mspace{1.0mu}\!F_1}(\alpha,\beta;\alpha;\,z)=(1-z)^{-\beta}$
and
\begin{gather}
\lim_{c\to\,2-l,\,l=1,2,\ldots}
{_2\mspace{1.0mu}\!F_1}(a,b;\,c;\,z)/{\Gamma(c)}=
\tag{$3.34$}\\
=
\dfrac{(a)_{l-1}(b)_{l-1}}{(l-1)!}z^{l-1}
{_2\mspace{1.0mu}\!F_1}(a+l-1,b+l-1;\,l;\,z),
\notag
\end{gather}
in the case
$a=1$, $b=\lambda_{sj}=-\,\omega-j+1-l$,  $c=2-l$,
from Eqs.(3.30) and (3.34) it follows
\begin{gather}
(R_{\mspace{1.0mu}\scriptscriptstyle 0}^{\mspace{1.0mu}\nu}
{\cal F})_{sj}=
M_{\epsilon}^{\,\omega+j}\,\Gamma{(-\,\omega-j)}
{_2\mspace{1.0mu}\!F_1}(l,-\,\omega-j;\,l;\,Z_{\epsilon})
\notag\\[2pt]
\phantom{%
(R_{\mspace{1.0mu}\scriptscriptstyle 0}^{\mspace{1.0mu}\nu}
{\cal F})_{sj}\,
}%
=
{\cal F}_{sj},
\text{ if $\nu_{sj}=-\,l,\ l\in\mathbb{N}_{\sscr{+}}$}\,,
\tag{$3.35$}
\end{gather}
i.e., the first relation in Eqs.(2.2).

Secondly,
the basic functions
$(R_{\mspace{1.0mu}\scriptscriptstyle 0}^{\mspace{1.0mu}\nu}
{\cal F})_{sj}\equiv$
$(R_{\mspace{1.0mu}\scriptscriptstyle 0}^{\mspace{1.0mu}\nu}
{\cal F})_{sj}(\omega;M_{\epsilon},A)$
of the self-consistently renormalized FAs obey
{\it the same recurrence relations}
as the basic functions
${\cal F}_{sj}\equiv{\cal F}_{sj}(\omega;M_{\epsilon},A)$
of convergent or dimensionally regularized FAs.
Really, if the recurrence relation,
see~\cite{
Bateman&Erdelyi1965}\
Ch.\;2.8, eq.42, i.e.,
\begin{gather}
(c-b-1)\,{_2\mspace{1.0mu}\!F_1}(a,b;\,c;\,z)+
b\,{_2\mspace{1.0mu}\!F_1}(a,b+1;\,c;\,z)-
\notag\\[2pt]
-
(c-1)\,{_2\mspace{1.0mu}\!F_1}(a,b;\,c-1;\,z)=0,
\tag{$3.36$}
\end{gather}
between contiguous Gauss hypergeometric functions
${_2\mspace{1.0mu}\!F_1}$
in the case
$a=1$, $b=\lambda_{sj}$, $c=2+\nu_{sj}$,
to multiply on the quantity
$M_{\epsilon}^{\,\omega+j}Z_{\epsilon}^{1+\nu_{sj}}
\Gamma(\lambda_{sj})/\Gamma(2+\nu_{sj})$
and to use of relations
$\nu_{s-2,j-1}=\nu_{sj}$,
$\lambda_{s-2,j-1}(\omega)=\lambda_{sj}(\omega)+1$,
and
$\nu_{s,j-1}=\nu_{sj}-1$,
$\lambda_{s,j-1}(\omega)=\lambda_{sj}(\omega)$,
then we obtain the following recurrence relations
\begin{gather}
M_{\epsilon}\,
(R_{\mspace{1.0mu}\scriptscriptstyle 0}^{\mspace{1.0mu}\nu}
{\cal F})_{s-2,j-1}-
A\,
(R_{\mspace{1.0mu}\scriptscriptstyle 0}^{\mspace{1.0mu}\nu}
{\cal F})_{s,j-1}+
\notag\\[3pt]
\hspace{20mm}
+\,
(\omega+j)\,
(R_{\mspace{1.0mu}\scriptscriptstyle 0}^{\mspace{1.0mu}\nu}
{\cal F})_{sj}=0,
\tag{$3.37$}
\end{gather}
between basic functions
$(R_{\mspace{1.0mu}\scriptscriptstyle 0}^{\mspace{1.0mu}\nu}
{\cal F})_{sj}\equiv$
$(R_{\mspace{1.0mu}\scriptscriptstyle 0}^{\mspace{1.0mu}\nu}
{\cal F})_{sj}(\omega;M_{\epsilon},A)$,
i.e., the second relation in Eqs.(2.1).

\smallskip
{\bf 3.7}
Transformation formulae
(see~\cite{Bateman&Erdelyi1965}\
Ch.\;2.1.4, eqs.22 and 23)
of
${_2\mspace{1.0mu}\!F_1}$
give rise to the representations:
\begin{gather}
(R_{\mspace{1.0mu}\scriptscriptstyle 0}^{\mspace{1.0mu}\nu}
{\cal F})_{sj}=
\dfrac
{(-1)\Gamma(\lambda_{sj}) A^{\nu_{sj}}}
{\Gamma(2+\nu_{sj}) M_{\epsilon}^{\,\lambda_{sj}-1} }
\left( \dfrac {Z_{\epsilon}} {Z_{\epsilon}-1} \right)
\notag\\[3pt]
\phantom{%
(R_{\mspace{1.0mu}\scriptscriptstyle 0}^{\mspace{1.0mu}\nu}
}%
\!\!\cdot
{_2\mspace{1.0mu}\!F_1} \left(1,\omega+j+1;\,2+\nu_{sj};
\dfrac {Z_{\epsilon}} {Z_{\epsilon}-1} \right),
\tag{$3.38$}\\[3pt]
(R_{\mspace{1.0mu}\scriptscriptstyle 0}^{\mspace{1.0mu}\nu}
{\cal F})_{sj}=
(M_{\epsilon}-A)^{\omega+j}
\dfrac {\Gamma(\lambda_{sj})} {\Gamma(2+\nu_{sj})}
Z_{\epsilon}^{\,1+\nu_{sj}}
\notag\\[3pt]
\phantom{%
(R_{\mspace{1.0mu}\scriptscriptstyle 0}^{\mspace{1.0mu}\nu}
}%
\!\!\cdot
{_2\mspace{1.0mu}\!F_1}(1+\nu_{sj},\omega+j+1;\,2+\nu_{sj};\,Z_{\epsilon}).
\tag{$3.39$}
\end{gather}
The equation (3.38) and the behaviour of
${_2\mspace{1.0mu}\!F_1}(a,b;\,c;\,z)$
in the vicinity
$z\to\,1_{-}$
to determine completely the asymptotic of basic functions
$(R_{\mspace{1.0mu}\scriptscriptstyle 0}^{\mspace{1.0mu}\nu}{\cal F})_{sj}$
for $A<0$ in the vicinity
$M_{\epsilon}\to\,0$, i.e., the chiral limit:
\begin{gather}
(R_{\mspace{1.0mu}\scriptscriptstyle 0}^{\mspace{1.0mu}\nu}
{\cal F})_{sj}
\overset{M_{\epsilon}\to 0}{\simeq}
\dfrac {(-1)\Gamma(\lambda_{sj}-1) A^{\nu_{sj}} }
{\Gamma(1+\nu_{sj})\,M_{\epsilon}^{\,\lambda_{sj}-1} },
\notag\\[3pt]
\phantom{%
(R_{\mspace{1.0mu}\scriptscriptstyle 0}^{\mspace{1.0mu}\nu}
{\cal F})_{sj}\
}%
\text{if $\nu_{sj} \ge 0$ and $\lambda_{sj}-1>0 $};
\notag\\[3pt]
(R_{\mspace{1.0mu}\scriptscriptstyle 0}^{\mspace{1.0mu}\nu}
{\cal F})_{sj}
\overset{M_{\epsilon}\to 0}{\simeq}
\dfrac {(-1)A^{\nu_{sj}}} {\Gamma(1+\nu_{sj}) }
\ln \left( 1- A/M_{\epsilon} \right),
\tag{$3.40$}\\[3pt]
\phantom{%
(R_{\mspace{1.0mu}\scriptscriptstyle 0}^{\mspace{1.0mu}\nu}
{\cal F})_{sj}\
}%
\text{if $\nu_{sj} \ge 0$ and $\lambda_{sj}-1=0 $};
\notag\\[3pt]
(R_{\mspace{1.0mu}\scriptscriptstyle 0}^{\mspace{1.0mu}\nu}
{\cal F})_{sj}
\overset{M_{\epsilon}\to 0}{\simeq}
\Gamma(-\,\omega -j)(-A)^{\omega +j},
\notag\\[3pt]
\phantom{%
(R_{\mspace{1.0mu}\scriptscriptstyle 0}^{\mspace{1.0mu}\nu}
{\cal F})_{sj}\
}%
\text{if $\nu_{sj}\ge 0$ and $\lambda_{sj}-1<0$ or $\nu_{sj}\leq -1$,}
\notag
\end{gather}
which is equivalent also to the asymptotic behaviour of
the basic functions in the case
$A\to -\infty$,\  $M_{\epsilon}\not=\,0$.
From Eqs.(3.30) follows four different series of
values for $\lambda_{sj}-1$:
\begin{gather}
\lambda_{sj}-1=
-\delta_n\delta_{\scriptscriptstyle |{\cal C}|}/2+
\notag\\[1pt]
\phantom{%
\lambda_{sj}-1
}%
+
(r_d-r_s)
+[(\delta_n\delta_{\scriptscriptstyle |{\cal C}|}+\delta_d-\delta_s)/2],
\tag{$3.41$}\\[1pt]
\lambda_{sj}-1=
(r_d-r_s)-1/2,\
\delta_n\delta_{\scriptscriptstyle |{\cal C}|}=1
\:\&\:\delta_s\geq\delta_d;
\notag\\[1pt]
\phantom{%
\lambda_{sj}-1
}%
=
(r_d-r_s)+1/2,\
\delta_n\delta_{\scriptscriptstyle |{\cal C}|}=1
\:\&\:\delta_d > \delta_s;
\notag\\[1pt]
\phantom{%
\lambda_{sj}-1
}%
=
(r_d-r_s),\
\phantom{+1/2\ \,}%
\delta_n\delta_{\scriptscriptstyle |{\cal C}|}=0
\:\&\:\delta_d\geq\delta_s;
\notag\\[1pt]
\phantom{%
\lambda_{sj}-1
}%
=(r_d-r_s)-1,\
\phantom{/2\,}%
\delta_n\delta_{\scriptscriptstyle |{\cal C}|}=0
\:\&\:\delta_s > \delta_d;
\tag{$3.42$}\\[1pt]
d=2r_d+\delta_d,\ s=2r_s+\delta_s,\ \
\delta_n,\:\delta_{\scriptscriptstyle |{\cal C}|},\:\delta_d,\:\delta_s=0,1.
\notag
\end{gather}
It is evident that Eq.(3.39) to present
a multiplicative realization of the subtraction procedure explicitly,
\begin{gather}
(R_{\mspace{1.0mu}\scriptscriptstyle 0}^{\mspace{1.0mu}\nu}
{\cal F})_{sj}\!:=
{\cal F}_{sj}-
(S_{\mspace{1.0mu}\scriptscriptstyle 0}^{\mspace{1.0mu}\nu}
{\cal F})_{sj}=
\notag\\[3pt]
\phantom{%
(R_{\mspace{1.0mu}\scriptscriptstyle 0}^{\mspace{1.0mu}\nu}
}%
\!\!=
{\cal F}_{sj}(\omega;M_{\epsilon},A)\,
({\Pi}_{\mspace{1.0mu}\scriptscriptstyle 0}^{\mspace{1.0mu}\nu}
{\cal F})_{sj}
(\omega;Z_{\epsilon}),
\notag\\[3pt]
({\Pi}_{\mspace{1.0mu}\scriptscriptstyle 0}^{\mspace{1.0mu}\nu}
{\cal F})_{sj}
(\omega;Z_{\epsilon})\!:=
\dfrac {(-\,\omega-j)_{1+\nu_{sj}}} {\Gamma(2+\nu_{sj})}
Z_{\epsilon}^{\,1+\nu_{sj}}
\tag{$3.43$}\\[3pt]
\phantom{%
({\Pi}_{\mspace{1.0mu}\scriptscriptstyle 0}^{\mspace{1.0mu}\nu}
}%
\!\!\cdot
{_2\mspace{1.0mu}\!F_1}(1+\nu_{sj},\omega+j+1;2+\nu_{sj};Z_{\epsilon}).
\notag
\end{gather}

\section{Homogeneous $\bf k$-polynomials
${\cal P}^{\scriptscriptstyle G}_{sj}(m,\alpha,k)$\\
of $\alpha$-parametric representation of FAs}

{\bf 4.1}
From Eq.(3.7) it is evident that
the basic functions
$(R_{\mspace{1.0mu}\scriptscriptstyle 0}^{\mspace{1.0mu}\nu}
{\cal F})_{sj}$
and the homogeneous $k$-polynomials
${\cal P}^{\scriptscriptstyle G}_{sj}(m,\alpha,k)$
in external momenta $k_e,\, e\in {\cal E}$, of degree $s-2j$,
$j=0,1,\ldots,[s/2]$,
are two closely related in pairs important universal ingrediants
of the SCR representation of FAs.
The latter are $\alpha$-images of the homogeneous $p$-polynomials
${\cal P}^{\scriptscriptstyle G}_{s}(m,p)$
in internal momenta $p_l,\, l\in {\cal L}$,
of degree $s$,
$s=0,1,\ldots, d^{\mspace{1.0mu}\scriptscriptstyle G}$,
appearing in the numerator polynomial
${\cal P}^{\scriptscriptstyle G}(m,p)$, see Eqs.(3.1)-(3.2).

Each monomial of
${\cal P}^{\scriptscriptstyle G}_{sj}(m,\alpha,k)$
is a product of $s-2j$ linear Kirchhoff forms
$Y_l(\alpha,k)\!:=\sum_{e\in {\cal E}}Y_{le}(\alpha)k_e$
and $j$ line-correlator functions
$X_{ll'}(\alpha),\,l,l'\in {\cal L}$, of a graph $G$.
The efficient algorithm of finding these expressions
from initial homogeneous $p$-polynomials
${\cal P}^{\scriptscriptstyle G}_s(m,p)$
in internal momenta $p_l,\, l\in {\cal L}$,
of degree $s=0,1,\ldots,d^{\mspace{1.0mu}\scriptscriptstyle G}$,
has been elaborated in~\cite{
Kucheryavy1974,
Kucheryavy1977,
Kucheryavy1979,
Kucheryavy1982}.
It resembles Wick relations between time-ordered and normal
products of boson fields in quantum field theory.
The main steps of this algorithm are as following.

${\bullet}$
The polynomials ${\cal P}^{\scriptscriptstyle G}_{s{\sscr0}}(m,\alpha,k)$
are determined as
\begin{gather}
{\cal P}_{s{\sscr0}}^{\scriptscriptstyle G}(m,\alpha,k)\!:=
{\cal P}_s^{\scriptscriptstyle G}(m,p)|_{p_l=Y_l(\alpha,k)},
\notag\\
j=0,\
s=0,1,\ldots, d^{\mspace{1.0mu}\scriptscriptstyle G},
\tag{$4.1$}
\end{gather}
i.e., by the straightforward substitution
$p_l\rightarrow Y_l(\alpha,k)$, $\forall l\in{\cal L}$,
in the polynomials
${\cal P}^{\scriptscriptstyle G}_s(m,p)$.

${\bullet}$
The polynomials ${\cal P}^{\scriptscriptstyle G}_{sj}(m,\alpha,k)$,
$j=1,\ldots, [s/2]$,
have the algebraic structure of quantities generated by the Wick
formula which represents a $T$-product of $s$ boson fields in terms
of some set of $N$-products of $s-2j$ boson fields with $j$ primitive
contractions.
In this case the linear Kirchhoff forms
$Y_l^{\sigma}(\alpha,k)$ and their primitive correlators
%

\begin{gather}
\underbrace{%
Y^{\sigma_1}_{l_1}
\cdots\,
Y^{\sigma_2}_{l_2}
}%
\!:=
(-1/2)X_{l_1l_2}(\alpha)g^{\sigma_1\sigma_2}\equiv
\notag\\
\phantom{%
Y^{\sigma_1}_{l_1}
\cdots\,
Y^{\sigma_2}_{l_2}\,
}%
\equiv
(-1/2)(\,{}_{\,l_1\,l_2}^{\sigma_1\sigma_2}).
\tag{$4.2$}
\end{gather}
play a role of boson fields and contractions, respectively.

\smallskip
{\bf 4.2}
As a result, we come to the following general formulae.
As far as homogeneous $p$-polynomials
${\cal P}^{\scriptscriptstyle G}_s(m,p)$
can be always represented as
\begin{gather}
{\cal P}^{\scriptscriptstyle G}_s(m,p)=
\sum\limits_{(i)\in G}a_s^{(i)}(m)\,
p_{l_1^{(i)}}^{\sigma_1^{(i)}}
p_{l_2^{(i)}}^{\sigma_2^{(i)}}
\cdots\,\,
p_{l_s^{(i)}}^{\sigma_s^{(i)}},\
\notag\\
l_a^{(i)}\in {\cal L},\ a={1,\ldots,s},
\tag{$4.3$}
\end{gather}
where the coefficients $a_s^{(i)}(m)$ are functions in masses
$m_l,\, l\in {\cal L}$,
it is sufficient to find the image of some
general monomial entering into the sum over
$(i)\in G$ in Eq.(4.3).
Calculation according to the above mentioned Wick type rule yields
\begin{gather}
p_{l_1^{(i)}}^{\sigma_1^{(i)}}
p_{l_2^{(i)}}^{\sigma_2^{(i)}}
\cdots\,\,
p_{l_s^{(i)}}^{\sigma_s^{(i)}}
\rightarrow
\sum_{j=0}^{[s/2]}
{\cal P}_{(l_1^{(i)}\cdots\,\,l_s^{(i)});j}
^{\mspace{1.0mu}\sigma_1^{(i)}\cdots\,\,\sigma_s^{(i)}}
(\alpha,k),
\notag\\
{\cal P}_{(l_1^{(i)}\cdots\,\,l_s^{(i)});j}
^{\mspace{1.0mu}\sigma_1^{(i)}\cdots\,\,\sigma_s^{(i)}}(\alpha,k)
=(-2)^{-j}
\notag\\
\hspace{10mm}
\cdot
\sum\limits_{d\in (1^{s-2j}2^j)}
{\cal P}_{(l_{d(1)}^{(i)}\cdots\,\,l_{d(s)}^{(i)});j}
^{ \mspace{1.0mu}\sigma_{d(1)}^{(i)}\cdots\,\,\sigma_{d(s)}^{(i)} }
(\alpha,k),
\notag\\
{\cal P}_{(l_{d(1)}^{(i)}\cdots\,\,l_{d(s)}^{(i)});j}
^{ \mspace{1.0mu}\sigma_{d(1)}^{(i)}\cdots\,\,\sigma_{d(s)}^{(i)} }
(\alpha,k)\!:=
\prod\limits^{s-2j}
Y_{l_{d(a)}^{(i)}}^{\mspace{1.0mu}\sigma_{d(a)}^{(i)}}(\alpha,k)
\notag\\
\hspace{15mm}
\cdot
\prod\limits^{j}
\bigl( X_{l_{d(b)}^{(i)}l_{d(c)}^{(i)}}(\alpha)
g^{\sigma_{d(b)}^{(i)}\sigma_{d(c)}^{(i)}} \bigr),
\tag{$4.4$}
\end{gather}
where the summation in the second Eq.(4.4)
is extended over all partitions
$d$ of $(l_1^{(i)},l_2^{(i)},\ldots,l_s^{(i)})$ according to
the Young scheme $(1^{s-2j}2^j)$.
Then the image of homogeneous
$p$-polynomials
${\cal P}^{\scriptscriptstyle G}_s(m,p)$ given by Eq.(4.3) is
\begin{align}
&
{\cal P}^{\scriptscriptstyle G}_s(m,p)
\rightarrow
\sum_{j=0}^{[s/2]}
{\cal P}^{\scriptscriptstyle G}_{sj}(m,\alpha,k),
\tag{$4.5$}\\
&
{\cal P}^{\scriptscriptstyle G}_{sj}(m,\alpha,k)=
\sum_{(i)\in G}a_s^{(i)}(m)
{\cal P}_{(l_1^{(i)}\cdots\,\,l_s^{(i)});j}
^{\mspace{1.0mu}\sigma_1^{(i)}\cdots\,\,\sigma_s^{(i)}}(\alpha,k).
\notag
\end{align}
In so doing, we are arrived to special
$j$-degree homogeneous polynomials in variables
$(\,{}_{\,l_1\,l_2}^{\sigma_1\sigma_2})$
involved in primitive correlators, see Eq.(4.2).
Polynomials of this type was introduced and named as hafnians
by Caianiello~\cite{
Caianiello1953,
Caianiello1973}
in the course of his QED investigations.
Hafnians are the counterparts of phaffians and closely connected with
permanents.
The simplest nontrivial hafnian
$(\,{}_{\,l_1\,l_2\,\,l_3\,\,l_4}^{\sigma_1\sigma_2\sigma_3\sigma_4})$
of 2-degree is given below in the last three lines of Eq.(4.6).

\smallskip
{\bf 4.3}
Taking into account the very important applied meaning of an algorithm
of constructing a family of homogeneous $k$-polynomials
${\cal P}^{\scriptscriptstyle G}_{sj}(m,\alpha,k)$
from initial $p$-polynomials
${\cal P}^{\scriptscriptstyle G}_s(m,p)$,
we give some examples:
\begin{align}
1
\hspace{-1mm}
&\rightarrow &&
\hspace{-3mm}
j\!=\!0\!:\! 1;
\notag\\[2pt]
p_l^{\sigma}
\hspace{-1mm}
&\rightarrow &&
\hspace{-3mm}
j\!=\!0\!:\!
Y_l^{\sigma}=:\![\,{}_{\,l}^{\sigma}];
\notag\\[2pt]
p_{l_1}^{\sigma_1}p_{l_2}^{\sigma_2}
\hspace{-1mm}
&\rightarrow &&
\hspace{-3mm}
j\!=\!0\!:\!
Y_{l_1}^{\sigma_1}Y_{l_2}^{\sigma_2}=:\!
[\,{}_{\,l_1\,l_2}^{\sigma_1\sigma_2}],
\notag\\[2pt]
{}&\phantom{\hspace{-1mm}\rightarrow} &&
\hspace{-3mm}
j\!=\!1\!:\! (-\tfrac{1}{2})
\{
X_{l_1l_2}g^{\sigma_1\sigma_2}=:\!
(\,{}_{\,l_1\,l_2}^{\sigma_1\sigma_2})
\};
\notag\\[2pt]
%
p_{l_1}^{\sigma_1}p_{l_2}^{\sigma_2}p_{l_3}^{\sigma_3}\!
\hspace{-1mm}
&\rightarrow &&
\hspace{-3mm}
j\!=\!0\!:
Y_{l_1}^{\sigma_1}Y_{l_2}^{\sigma_2}Y_{l_3}^{\sigma_3}=:\!
[{}_{\,l_1\,l_2\,\,l_3}^{\sigma_1\sigma_2\sigma_3}],
\notag\\[2pt]
{}&\phantom{\hspace{-1mm}\rightarrow} &&
\hspace{-3mm}
j\!=\!1\!:\! (-\tfrac{1}{2})
\{\,
(\,{}_{\,l_1\,l_2}^{\sigma_1\sigma_2})[\,{}_{\,l_3}^{\sigma_3}]
{\scriptstyle +}
\notag\\[2pt]
{}&\phantom{\hspace{-1mm}\rightarrow} &&
\hspace{-3mm}
\phantom{ j\!=\!1\!:\!}
{\scriptstyle +}
(\,{}_{\,l_1\,l_3}^{\sigma_1\sigma_3})[\,{}_{\,l_2}^{\sigma_2}]
\,{\scriptstyle +}\,
(\,{}_{\,l_2\,l_3}^{\sigma_2\sigma_3})[\,{}_{\,l_1}^{\sigma_1}]\,\};
\notag\\[2pt]
p_{l_1}^{\sigma_1}p_{l_2}^{\sigma_2}p_{l_3}^{\sigma_3}p_{l_4}^{\sigma_4}
\hspace{-1mm}
{}&\rightarrow &&
\hspace{-3mm}
j\!=\!0\!:\!
Y_{l_1}^{\sigma_1}Y_{l_2}^{\sigma_2}Y_{l_3}^{\sigma_3}
Y_{l_4}^{\sigma_4}\!=:\!
[\,{}_{\,l_1\,l_2\,\,l_3\,l_4}^{\sigma_1\sigma_2\sigma_3\sigma_4}],
\notag\\[3pt]
{}&\phantom{\hspace{-1mm}\rightarrow} &&
\hspace{-3mm}
j\!=\!1\!:\! (-\tfrac{1}{2})
\{\,
(\,{}_{\,l_1\,l_2}^{\sigma_1\sigma_2})
[\,{}_{\,l_3\,l_4}^{\sigma_3\sigma_4}]\,
{\scriptstyle +}
\notag\\[3pt]
{}&\phantom{\hspace{-1mm}\rightarrow} &&
\hspace{-3mm}
\phantom{ j\!=\!1\!:\!}
{\scriptstyle +}\,
(\,{}_{\,l_1\,l_3}^{\sigma_1\sigma_3})
[\,{}_{\,l_2\,l_4}^{\sigma_2\sigma_4}]
\,{\scriptstyle +}\,
(\,{}_{\,l_1\,l_4}^{\sigma_1\sigma_4})
[\,{}_{\,l_2\,l_3}^{\sigma_2\sigma_3}]
\notag\\[3pt]
{}&\phantom{\hspace{-1mm}\rightarrow} &&
\hspace{-3mm}
\phantom{ j\!=\!1\!:\!}
{\scriptstyle +}\,
(\,{}_{\,l_2\,l_3}^{\sigma_2\sigma_3})
[\,{}_{\,l_1\,l_4}^{\sigma_1\sigma_4}]
\,{\scriptstyle +}\,
(\,{}_{\,l_2\,l_4}^{\sigma_2\sigma_4})
[\,{}_{\,l_1\,l_3}^{\sigma_1\sigma_3}]
\notag\\[3pt]
{}&\phantom{\hspace{-1mm}\rightarrow} &&
\hspace{-3mm}
\phantom{ j\!=\!1\!:\!}
{\scriptstyle +}\,
(\,{}_{\,l_3\,l_4}^{\sigma_3\sigma_4})
[\,{}_{\,l_1\,l_2}^{\sigma_1\sigma_2}]\,
\},
\tag{$4.6$}\\[3pt]
{}&\phantom{\hspace{-1mm}\rightarrow} &&
\hspace{-3mm}
j\!=\!2\!:\! (-\tfrac{1}{2})^{\sscr 2}
\{\,
(\,{}_{\,l_1\,l_2}^{\sigma_1\sigma_2})
(\,{}_{\,l_3\,l_4}^{\sigma_3\sigma_4})
\,{\scriptstyle +}
\notag\\[3pt]
{}&\phantom{\hspace{-1mm}\rightarrow} &&
\hspace{-3mm}
\phantom{ j\!=\!2\!:\!}
{\scriptstyle +}\,
(\,{}_{\,l_1\,l_3}^{\sigma_1\sigma_3})
(\,{}_{\,l_2\,l_4}^{\sigma_2\sigma_4})
{\scriptstyle +}
(\,{}_{\,l_1\,l_4}^{\sigma_1\sigma_4})
(\,{}_{\,l_2\,l_3}^{\sigma_2\sigma_3})\}
\notag\\[3pt]
{}&\phantom{\hspace{-1mm}\rightarrow} &&
\hspace{-3mm}
\phantom{ j\!=\!2\!:\!}
=:\!
(-\tfrac{1}{2})^{\sscr 2}
(\,{}_{\,l_1\,l_2\,\,l_3\,\,l_4}^{\sigma_1\sigma_2\sigma_3\sigma_4}).\,
\notag
\end{align}

\section{Parametric functions of FAs }

{\bf 5.1}
Now, let us formulate an algorithm of finding the parametric
functions
\begin{equation*}
\Delta(\alpha),\quad A(\alpha,k),\quad Y_l(\alpha,k),\quad
X_{ll'}(\alpha),\ l,l'\in {\cal L},
\end{equation*}
of Feynman amplitudes.
Of course, it is to be mentioned that
we can in principle use any one of the available approaches.
Contributions to this subject have been made by many authors.
We point out here the very incomplete list of quotes,
namely papers by
Chisholm~\cite{
Chisholm1952},
Nambu~\cite{
Nambu1957},
Symanzik~\cite{
Symanzik1958},
Nakanishi~\cite{
Nakanishi1961},
Schimamoto~\cite{
Schimamoto1962},
Bjorken and Wu~\cite{
Bjorken&Wu1963},
Peres~\cite{
Peres1965},
Lam and Lebrun~\cite{
Lam&Lebrun1969},
Stepanov~\cite{
Stepanov1970},
Liu and Chow~\cite{
Liu&Chow1970},
Cvitanovic and Kinoshita~\cite{
Cvitanovic&Kinoshita1974},
and books by
Todorov~\cite{
Todorov1966},
Speer~\cite{
Speer1969},
Nakanishi~\cite{
Nakanishi1971},
Zavialov~\cite{
Zavialov1979},
Smirnov~\cite{
Smirnov1990},
in which many other citations can be found.
Nevertheless, our algorithm seems to be very simple,
but universal enough. It is named by the author~\cite{
Kucheryavy1972,
Kucheryavy1972a,
Kucheryavy1977}
as {\it circuit-path} algorithm.

{\bf 5.2}
Suppose we have connected graph
$G({\cal V},{\cal L}\cup{\cal E})$
with sets of vertices, ${\cal V}$, of internal lines, ${\cal L}$,
of external lines, ${\cal E}$, and with a certain relation of
incidence between ${\cal V}$ and $\Lambda\equiv{\cal L}\cup{\cal E}$,
described by an oriented incidence matrix
$e_{il}\equiv [e_{\sscr{\cal V}\Lambda}]_{il}=0,\,\pm 1$,
$v_i\in{\cal V},\ l\in\Lambda$.
In particular, $e_{il}=0$, if the line $l$ is nonincident to the
vertex $v_i$; $e_{il}=1$, if the line $l$ is outgoing from the
vertex $v_i$; $e_{il}=-1$, if the line $l$ is incoming to the
vertex $v_i$.
The fact that the set of all lines $\Lambda$ is separated from
the beginning into two mutually disjoint subsets ${\cal L}$ and
${\cal E}$ (incident properties of which are different)
is very important
both from algorithmic point of view
and from potential possibilities.
In so doing, we need not replace here the set of external lines
(incident to some one vertex) by some effective line,
or assign the same orientation to all external lines,
as is usually done.
Therefore,
we can set the task of constructing parametric functions
of whole graph via parametric functions of its subgraphs.
As a result, circuit-path approach is naturally arose and the
recursive structure of parametric functions of FAs
has been obtained~\cite{
Kucheryavy1972a,
Kucheryavy1972b}.

{\bf 5.3}
The set of external lines, $\cal E$, induces the single-valued
decomposition of the set of all vertices, $\cal V$,
into the subset of external vertices, ${\cal V}^{ext}$,
and the subset of internal vertices, ${\cal V}^{int}$.
The set of internal lines, $\cal L$, can be always decompose
(as a rule in more than one way) into two mutually disjoint subsets,
$\cal M$ and $\cal N$, which determine some {\it skeleton tree} and
corresponding to it {\it co-tree} subgraphs of the graph $G$.
So, we have the following decomposition of the set
$\Lambda={\cal E}\cup{\cal N}\cup{\cal M}$
of all lines of the graph $G$ into mutually disjoint subsets,
${\cal E}$, ${\cal N}$, and ${\cal M}$.
Then circuit-path algorithm requires the following steps:

$\bullet$
Let us choose a subset ${\cal N}\subset{\cal L}$ such that the subgraph
$G({\cal V},{\cal M}\cup{\cal E})$,
where ${\cal M}\!:={\cal L}/{\cal N}$,
is a {\it skeleton tree} type graph and the subgraph
$G({\cal V},{\cal N}\cup{\cal E})$
is a {\it co-tree} type graph.
It is clear that this choice is ambiguous.
It is shown in~\cite{
Kucheryavy1972},
however, that the parametric functions
are independent of any choice of ${\cal N}$.

$\bullet$
Let us choose a vertex $v_j\in {\cal V}$ which will be refered to as
a {\it basis vertex} (or {\it reference vertex}, or {\it zero point}).
It is clear that this choice is also ambiguous.
But it is shown in~\cite{
Kucheryavy1972},
that the parametric functions are again
independent of any given choice of $v_j$.
From the viewpoint of practical calculations it seems reasonable
to choose the basis vertex $v_j$ as such a vertex to which
the largest number of external lines of the graph are incident.

$\bullet$
The choice of ${\cal N}\subset{\cal L}$ and the basis vertex
$v_j$ uniquely defines notions of basis circuits
$C(n),\,n\in {\cal N}$, and basis paths $P(j|e),\,e\in {\cal E}$, namely.

The {\it basis circuit} $C(n)$ generated by the line $n\in {\cal N}$
is a union of the line $n$ with the subset ${\cal M}(n)\subset{\cal M}$
which forms a chain in ${\cal M}$ between vertices incident to
the line $n$, i.e. $C(n)\!:=\{n\}\cup{\cal M}(n)$.
The orientation in the circuit $C(n)$
is defined by the orientation of the line $n\in {\cal N}$.

The {\it basis path} $P(j|e)$ generated by the line $e\in {\cal E}$ and
the basis vertex $v_j$ is a union of the line $e$ with the subset
${\cal M}(j|e)\subset{\cal M}$ which forms a chain in ${\cal M}$ between
a vertex incident to the line $e\in {\cal E}$ and the basis vertex $v_j$,
i.e. $P(j|e)\!:=\{e\}\cup{\cal M}(j|e)$.
The orientation in the path $P(j|e)$
is defined by the orientation of the line $e\in {\cal E}$.

$\bullet$
By analogy with the incidence matrix $e_{\sscr {\cal V}\Lambda}$,
that is more precisely can be refered to as
the {\it vertex-line} incidence matrix,
one introduces topologically the {\it line-circuit}
$e_{\sscr \Lambda{\cal N}}$~\cite{
Liu&Chow1970,
Cvitanovic&Kinoshita1974,
Nakanishi1971,
Kucheryavy1972,
Kucheryavy1972a},
and the {\it line-path}
$e_{\sscr \Lambda{\cal E}}(j)$~\cite{
Kucheryavy1972,
Kucheryavy1972a}
incidence matrices, namely:
\begin{gather}
[e_{\sscr\Lambda\cal N}]_{ln}\!:=
\left\{
\begin{array}{rl}
0,& l\notin C(n),\\[2pt]
\pm1,& l\in C(n);
\end{array}
\right.
\notag\\[3pt]
[e_{\sscr\Lambda\cal E}(j)]_{le}\!:=
\left\{
\begin{array}{rl}
0,& l\notin P(j|e),\\[3pt]
\pm1,& l\in P(j|e).
\end{array}
\right.
\tag{$5.1$}
\end{gather}
Here the plus or minus sign depends on whether the orientation of the
line $l\in\Lambda$ coincide or not with the orientation of the circuit
$C(n)$ for $e_{\sscr\Lambda\cal N}$ or the path
$P(j|e)$ for $e_{\sscr\Lambda{\cal E}}(j)$.
As a result,
the column-vector
$p_{\sscr\Lambda}$
of all momenta
$p_l, l\in\Lambda$,
and submatrices of
$e_{\sscr\Lambda\cal N}$ and
$e_{\sscr\Lambda{\cal E}}(j)$,
the rows of which are associated with the partition
$\Lambda={\cal E}\cup{\cal N}\cup{\cal M}$,
can be represented as follows~\cite{
Kucheryavy1972,
Kucheryavy1972a}:
\begin{gather}
p_{\sscr\Lambda}=p^{\,ext}_{\sscr\Lambda}+p^{\,int}_{\sscr\Lambda},\
p^{\,ext}_{\sscr\Lambda}=e_{\sscr\Lambda{\cal E}}(j)k_{\sscr{\cal E}},\
p^{\,int}_{\sscr\Lambda}=e_{\sscr\Lambda{\cal N}}p_{\sscr{\cal N}};
\notag\\[1pt]
e_{\sscr\cal E\cal E}(j)=1_{\sscr\cal E\cal E},\
e_{\sscr\cal N\cal E}(j|{\scriptstyle\cal N})=0_{\sscr\cal N\cal E},
\notag\\[2pt]
\phantom{%
e_{\sscr\cal E\cal E}(j)=1_{\sscr\cal E\cal E},\ \
}%
e_{\sscr\cal M\cal E}(j|{\scriptstyle\cal N})=
-
e^{-1}_{\{\sscr{\cal V}/j\}\cal M}e_{\{\sscr{\cal V}/j\}\cal E};
\notag\\[-3pt]
e_{\sscr\cal E\cal N}=0_{\sscr\cal E\cal N},\ \ \ \
e_{\sscr\cal N\cal N}=1_{\sscr\cal N\cal N},
\tag{$5.2$}\\ 
\phantom{%
e_{\sscr\cal E\cal E}(j)=1_{\sscr\cal E\cal E},\ \
}%
e_{\sscr\cal M\cal N}=
-
e^{-1}_{\{\sscr{\cal V}/j\}\cal M}e_{\{\sscr{\cal V}/j\}\cal N}.
\notag
\end{gather}
From now on,
$k_{\sscr{\cal E}}$ and
$p_{\sscr{\cal N}}$
are the column-vectors of external momenta
$k_e, e\in{\cal E}$, and
independent integration momenta
$p_n, n\in{\cal N}$,
respectively;
$0_{\sscr\cal A\cal B}$ is the
$|\cal A|\times|\cal B|$-rectangular
matrix of zeros, and
$1_{\sscr\cal A\cal A}$
is the $|\cal A|$-dimensional unit matrix.
Matrices
$e_{\{\sscr{\cal V}/j\}\cal E}$,\
$e_{\{\sscr{\cal V}/j\}\cal N}$,\
$e_{\{\sscr{\cal V}/j\}\cal M}$,
are submatrices of
$e_{\sscr{\cal V}\Lambda}$.
Their rows are defined by the set
$({\cal V}/v_j)\subset{\cal V}$,
and their columns are defined by the subsets
${\cal E}$,\ ${\cal N}$,\ ${\cal M}$,
respectively.
The $(|{\cal V}|-1)$-dimensional square matrix
$e_{\{\sscr{\cal V}/j\}\cal M}$
is nonsingular, and
$\det[e_{\{\sscr{\cal V}/j\}\cal M}]=\pm 1$.
In submatrices of the second and third lines of Eqs.(5.2),
the subset ${\cal N}$ is pointed out explicitly,
because of
$e_{\sscr\cal N'\cal E}(j|{\scriptstyle\cal N})\not=0_{\sscr\cal N'\cal E}$,
and
$e_{\sscr\cal M'\cal E}(j|{\scriptstyle\cal N})\not=
e_{\sscr\cal M\cal E}(j|{\scriptstyle\cal N})$
if ${\cal N}'\not={\cal N}$,
${\cal L}={\cal N}\cup{\cal M}={\cal N}'\cup{\cal M}'$,
but
$e_{\sscr\cal E\cal E}(j|{\scriptstyle\cal N})=
e_{\sscr\cal E\cal E}(j|{\scriptstyle\cal N'})=1_{\sscr\cal E\cal E}$.

$\bullet$
There exist the following very important ``orthogonality''
relations~\cite{
Bryant1967,
Kucheryavy1972,
Kucheryavy1972a}:
\begin{gather}
e_{\sscr{\cal V}\Lambda}e_{\sscr\Lambda{\cal N}}=
e_{\sscr{\cal V}\cal L}e_{\sscr\cal L\cal N}=0_{\sscr\cal V\cal N},
\notag\\
e_{\{\sscr{\cal V}/j\}\Lambda}e_{\sscr\Lambda\cal N}=
e_{\{\sscr{\cal V}/j\}\cal L}e_{\sscr\cal L\cal N}=
0_{\{\sscr{\cal V}/j\}\cal N},
\notag\\[2pt]
[e_{\sscr{\cal V}\Lambda}e_{\sscr\Lambda\cal E}(j)]_{ie}=
\delta_{ij}[e({\scriptstyle\cal V}^{\ast})_{\sscr\cal E}]_e,
\notag\\[2pt]
e_{\{\sscr{\cal V}/j\}\Lambda}e_{\sscr\Lambda\cal E}(j)=
0_{\{\sscr{\cal V}/j\}\cal E},
\tag{$5.3$}
\end{gather}
where $e({\scriptstyle\cal V}^{\ast})_{\sscr\cal E}$ is the vertex-line
incidence matrix of the {\it ``star''} type graph
$G^{\ast}\!:=<{\cal V}^{\ast}, {\cal E}>$
with the one vertex ${\cal V}^{\ast}$
and the set of external lines ${\cal E}$ of the graph $G$.
The graph
$G^{\ast}\!:=<{\cal V}^{\ast}, {\cal E}>$
is a result of shrinking of all vertices $v_i\in{\cal V}$,
and all internal lines $l\in{\cal L}$,
of the graph $G$ to the single {\it ``black-hole''} vertex
${\cal V}^{\ast}$.

$\bullet$
On assigning to every internal line $l\in {\cal L}$ the parameter
$\alpha_l\ge 0$,
we define the
{\it circuit}
$C_{\sscr\cal N\cal N}(\alpha)$,
{\it path}
$E_{\sscr\cal E\cal E}(j|\alpha)$,
and {\it path-circuit}
$\Pi_{\sscr\cal E\cal N}(j|\alpha)$
matrices~\cite{
Kucheryavy1972,
Kucheryavy1972a},
according to:
\begin{gather}
[C_{\sscr\cal N\cal N}(\alpha)]_{nn'}\!:=
[e^{\sscr T}_{\sscr\cal L\cal N}
\alpha_{\sscr\cal L\cal L}
e_{\sscr\cal L\cal N}]_{nn'}=
\notag\\ 
\hspace{20.0mm}
=
\pm
\sum_{l\in C(n)\cap C(n')}\,\alpha_l,
\notag\\ 
[E_{\sscr\cal E\cal E}(j|\alpha)]_{ee'}\!:=
[e^{\sscr T}_{\sscr\cal L\cal E}(j)
\alpha_{\sscr\cal L\cal L}
e_{\sscr\cal L\cal E}(j)]_{ee'}=
\notag\\ 
\hspace{20.5mm}
=
\pm
\sum_{l\in P(j|e)\cap P(j|e')}\,\alpha_l,
\notag\\ 
[\Pi_{\sscr\cal E\cal N}(j|\alpha)]_{en}\!:=
[e^{\sscr T}_{\sscr\cal L\cal E}(j)
\alpha_{\sscr\cal L\cal L}
e_{\sscr\cal L\cal N}]_{en}=
\notag\\ 
\hspace{21.0mm}
=
\pm
\sum_{l\in P(j|e)\cap C(n)}\alpha_l.
\tag{$5.4$}
\end{gather}
Here the plus or minus sign depends on the mutual orientations of the sets,
over which the summation is performed, on their intersection.
The plus sign corresponds to the case of coinciding orientations.
It is clear that the explicit form of these matrices
in any given case
can be easily obtained by inspecting the graph.
From now on
$\alpha_{\sscr\cal L\cal L}$
is the diagonal $|\cal L|$-dimensional matrix, i.e.,
$[\alpha_{\sscr\cal L\cal L}]_{ll'}=\alpha_l\,\delta_{ll'}$.

$\bullet$
The parametric functions
are derived by means of use the following matrices~\cite{
Kucheryavy1972,
Kucheryavy1972b}:
\begin{gather}
A_{\sscr\cal E\cal E}(j|\alpha)\!:=
E_{\sscr\cal E\cal E}(j|\alpha)-
\Pi_{\sscr\cal E\cal N}(j|\alpha)
C^{-1}_{\sscr\cal N\cal N}(\alpha)
\Pi^{\,\sscr T}_{\sscr\cal E\cal N}(j|\alpha),
\notag\\ 
Y_{\sscr\cal L\cal E}(j|\alpha)\!:=
e_{\sscr\cal L\cal E}(j)-
e_{\sscr\cal L\cal N}
C^{-1}_{\sscr\cal N\cal N}(\alpha)
\Pi^{\,\sscr T}_{\sscr\cal E\cal N}(j|\alpha),
\notag\\ 
X_{\sscr\cal L\cal L}(\alpha)\!:=
e_{\sscr\cal L\cal N}
C^{-1}_{\sscr\cal N\cal N}(\alpha)
e^{\,\sscr T}_{\sscr\cal L\cal N},
\notag\\ 
\Delta(\alpha)\!:=\det C_{\sscr\cal N\cal N}(\alpha).
\tag{$5.5$}
\end{gather}
So,
the quadratic
$A(\alpha,k)$
and linear
$Y_l(\alpha,k)$, $l\in{\cal L}$,
Kirchhoff forms in external momenta $k_e$, $e\in{\cal E}$,
and the line-correlator functions
$X_{ll'}(\alpha)$, $l,l'\in{\cal L}$,
are defined as~\cite{
Kucheryavy1972,
Kucheryavy1972b}:
\begin{gather}
A(\alpha,k)\!:=
(k_{\sscr\cal E}^{\sscr T}\cdot
A_{\sscr\cal E\cal E}(j|\alpha)
k_{\sscr\cal E})=
\notag\\[3pt]
\phantom{%
A(\alpha,k)\!:\!
}%
=
{\textstyle\sum\nolimits}_{e,e'\in {\cal E}}
[A_{\sscr\cal E\cal E}(j|\alpha)]_{ee'}
(k_e\cdot k_{e'}),
\notag\\[3pt]
Y_l(\alpha,k)\!:=
Y_{\sscr l\cal E}(j|\alpha)
k_{\sscr\cal E}=
{\textstyle\sum\nolimits}_{e\in {\cal E}}
[Y_{\sscr l{\cal E}}(j|\alpha)]_{e}k_e,\quad
\notag\\[3pt]
Y_{\sscr l{\cal E}}(j|\alpha)=
e_{\sscr l{\cal E}}(j)-
e_{\sscr l{\cal N}}
C^{-1}_{\sscr\cal N\cal N}(\alpha)
\Pi^{\,\sscr T}_{\sscr\cal E\cal N}(j|\alpha),
\notag\\[3pt]
X_{ll'}(\alpha)=
e_{\sscr l\cal N}
C^{-1}_{\sscr\cal N\cal N}(\alpha)
e^{\,\sscr T}_{\sscr l'\cal N}.
\tag{$5.6$}
\end{gather}
Here
$e_{\sscr l{\cal N}}$  and
$e_{\sscr l{\cal E}}(j)$
are row-vectors of matrices (5.1)-(5.2) for the line $l\in{\cal L}$.

{\bf 5.4}
It should be mentioned that functions
$\Delta(\alpha)$ and $A(\alpha,k)$
do not depend on the orientation of internal lines.
However, when the orientation of the line $l$ is changed,
parametric functions
$Y_l(\alpha,k)$ and $X_{ll'}(\alpha)$
are reversed their sign.

It is also useful to represent quantities
$A(\alpha,k)$ and
$Y_{\sscr\cal L}(\alpha,k)$
in a form exhibiting a special role of matrices
$X_{\sscr\cal L\cal L}(\alpha)$ and
$X_{\sscr\cal N\cal N}(\alpha)$~\cite{
Cvitanovic&Kinoshita1974,
Kucheryavy1972}:
\begin{gather}
A(\alpha,k)=
\notag\\[1pt]
=
\bigl(
p^{\,ext}_{\sscr\cal L}(k)^{\,\sscr T}\cdot
[\,\alpha_{\sscr\cal L\cal L}-
\alpha_{\sscr\cal L\cal L}
X_{\sscr\cal L\cal L}(\alpha)
\alpha_{\sscr\cal L\cal L}\,]\,
p^{\,ext}_{\sscr\cal L}(k) \bigr),
\notag\\[1pt]
p^{\,ext}_{\sscr\cal L}(k)=
e_{\sscr\cal L\cal E}(j)
k_{\sscr\cal E},\ \
p^{ext}_{\sscr\cal E}(k)=k_{\sscr\cal E},\ \
p^{ext}_{\sscr\cal N}(k)=0_{\sscr\cal N},
\notag\\[2pt]
Y_{\sscr\cal L}(\alpha,k)=
[\,1_{\sscr\cal L\cal L}-
X_{\sscr\cal L\cal L}(\alpha)
\alpha_{\sscr\cal L\cal L}\,]\,
p^{\,ext}_{\sscr\cal L}(k)=
\notag\\[2pt]
=
p^{\,ext}_{\sscr\cal L}(k)-
Y^{\,int}_{\sscr\cal L}(\alpha,k),
\notag\\[1pt]
Y^{\,int}_{\sscr\cal L}(\alpha,k)\!:=
X_{\sscr\cal L\cal L}(\alpha)
\alpha_{\sscr\cal L\cal L}\,
p^{\,ext}_{\sscr\cal L}(k),
\notag\\[2pt]
X_{\sscr\cal L\cal L}(\alpha)\!:=
e_{\sscr\cal L\cal N}
X_{\sscr\cal N\cal N}(\alpha)
e^{\,\sscr T}_{\sscr\cal L\cal N},
\notag\\[2pt]
X_{\sscr\cal N\cal N}(\alpha)\!:=
C^{-1}_{\sscr\cal N\cal N}(\alpha),
\tag{$5.7$}
\end{gather}
where $k_{\sscr\cal E}$ is the column-vector of external momenta
$k_e$, $e\in{\cal E}$.
There hold the following homogeneous properties:
\begin{gather}
\Delta(\rho\alpha)=\rho^{\,\sscr|\cal C|}\Delta(\alpha),\quad
X_{ll'}(\rho\alpha)=\rho^{-1}X_{ll'}(\alpha),
\notag\\[2pt]
A(\rho\alpha,\tau k)=\rho\tau^2 A(\alpha,k),\quad
Y_l(\rho\alpha,\tau k)=\tau Y_l(\alpha,k),
\notag\\[2pt]
{\cal P}^{\scriptscriptstyle G}_{sj}(m,\rho\alpha,\tau k)=
\rho^{-j}\tau^{s-2j}\,
{\cal P}^{\scriptscriptstyle G}_{sj}(m,\alpha,k).
\tag{$5.8$}
\end{gather}

{\bf 5.5}
Now we exhibit some important properties of $\alpha$-parametric
functions~\cite{
Kucheryavy2004}.
Let us introduce quantities,
\begin{gather}
K^r_{\sscr\cal L\cal L}\!:=
X_{\sscr\cal L\cal L}
\alpha_{\sscr\cal L\cal L},\ \
K^l_{\sscr\cal L\cal L}\!:=
\alpha_{\sscr\cal L\cal L}
X_{\sscr\cal L\cal L},\ \
\notag\\[2pt]
L^i_{\sscr\cal L\cal L}\!:=
1_{\sscr\cal L\cal L}-K^i_{\sscr\cal L\cal L},\ \
i=r,\,l.    
\tag{$5.9$}
\end{gather}
Using Eqs.(5.3)-(5.5), we find that matrices
$K^i_{\sscr\cal L\cal L}(\alpha)$ and
$L^i_{\sscr\cal L\cal L}(\alpha)$
are projectors with next properties:
\begin{gather}
K^i_{\sscr\cal L\cal L}
K^i_{\sscr\cal L\cal L}=
K^i_{\sscr\cal L\cal L},\ \
%
L^i_{\sscr\cal L\cal L}
L^i_{\sscr\cal L\cal L}=
L^i_{\sscr\cal L\cal L},\ \
i=r,l,
\notag\\[2pt]
K^i_{\sscr\cal L\cal L}
L^i_{\sscr\cal L\cal L}=
0_{\sscr\cal L\cal L},\ \
K^l_{\sscr\cal L\cal L}
\alpha_{\sscr\cal L\cal L}
L^r_{\sscr\cal L\cal L}=
0_{\sscr\cal L\cal L},\ \ \
\tag{$5.10$}
\end{gather}
From Eqs.(5.10), we get some
relations between products of
$X_{\sscr\cal L\cal L}$,
$\alpha_{\sscr\cal L\cal L}$, and
$Y_{\sscr\cal L\cal E}$:
\begin{gather}
\bigl(
X_{\sscr\cal L\cal L}
\alpha_{\sscr\cal L\cal L}
\bigr)^m
X_{\sscr\cal L\cal L}=
X_{\sscr\cal L\cal L}
\alpha_{\sscr\cal L\cal L}
X_{\sscr\cal L\cal L}=
X_{\sscr\cal L\cal L},\ \
\notag\\[2pt]
\bigl(
L^r_{\sscr\cal L\cal L}
\bigr)^m
X_{\sscr\cal L\cal L}=
0_{\sscr\cal L\cal L},
\notag\\[2pt]
\bigl(
X_{\sscr\cal L\cal L}
\alpha_{\sscr\cal L\cal L}
\bigr)^m
Y_{\sscr\cal L\cal E}\,\,=\,
X_{\sscr\cal L\cal L}
\alpha_{\sscr\cal L\cal L}
Y_{\sscr\cal L\cal E}\,=\,
0_{\sscr\cal L\cal E},\ \
\tag{$5.11$}\\[2pt]
\,\bigl(
L^r_{\sscr\cal L\cal L}
\bigr)^m
Y_{\sscr\cal L\cal E}\,=
Y_{\sscr\cal L\cal E},
\notag\\[2pt]
{\rm Tr}\bigl[\bigl( K^i_{\sscr\cal L\cal L} \bigr)^m\bigr]=
{\rm Tr}\bigl[ K^i_{\sscr\cal L\cal L} \bigr]=
{\textstyle\sum\nolimits}_{l\in{\cal L}}\,\alpha_lX_{ll}(\alpha)=
|{\cal N}|,
\notag\\[2pt]
{\rm Tr}\bigl[\bigl( L^i_{\sscr\cal L\cal L} \bigr)^m\bigr]\,=
{\rm Tr}\bigl[ L^i_{\sscr\cal L\cal L} \bigr]=
|{\cal M}|,\quad i=r,l,
\notag
\end{gather}
and the following relations between quadratic $A(\alpha,k)$ and
linear $Y_l(\alpha,k)$, $l\in\cal L$,
Kirchhoff forms:
\begin{align}
A(\alpha,k)&=
(Y_{\sscr\cal L}^{\,\sscr T}\cdot
\alpha_{\sscr\cal L\cal L}
p^{\,ext}_{\sscr\cal L})=
(p^{\,ext\,\sscr T}_{\sscr\cal L}
\cdot
\alpha_{\sscr\cal L\cal L}
Y_{\sscr\cal L})\equiv
\notag\\[3pt]
{}&\equiv
{\textstyle\sum\nolimits}_{l\in{\cal L}}\,
\alpha_l(p_l^{\,ext}(k)\cdot Y_l(\alpha,k))=
\tag{$5.12$}\\[3pt]
{}&=
(Y^{\,\sscr T}_{\sscr\cal L}\cdot
\alpha_{\sscr\cal L\cal L}
Y_{\sscr\cal L})\equiv
{\textstyle\sum\nolimits}_{l\in{\cal L}}
\alpha_l Y_l^2(\alpha,k).
\notag
\end{align}
There exist also the following relations,
\begin{gather}
e_{\sscr\cal V\cal E}k_{\sscr\cal E}+
e_{\sscr\cal V\cal L}
Y_{\sscr\cal L}(\alpha,k)=
0_{\sscr\cal V},\ \
e^{\,\sscr T}_{\sscr\cal L\cal N}
\alpha_{\sscr\cal L\cal L}
Y_{\sscr\cal L}(\alpha,k)=
0_{\sscr\cal N},
\notag\\[2pt]
e_{\sscr\cal V\cal L}
X_{\sscr\cal L\cal L}(\alpha)=
0_{\sscr\cal V\cal L},
\notag\\[2pt]
K^r_{\sscr\cal L\cal L}
e_{\sscr\cal L\cal E}(j)=
-e_{\sscr\cal L\cal N}
Y_{\sscr\cal N\cal E}(j|\alpha)=
e_{\sscr\cal L\cal N}
K^r_{\sscr\cal N\cal L}
e_{\sscr\cal L\cal E}(j),
\notag\\[2pt]
K^r_{\sscr\cal L\cal L}
e_{\sscr\cal L\cal N}
=e_{\sscr\cal L\cal N},
\tag{$5.13$}\\[2pt]
\bigl(
Y^{\,int\,\sscr T}_{\sscr\cal L}\cdot
\alpha_{\sscr\cal L\cal L}
Y_{\sscr\cal L}
\bigr)=
\notag\\[2pt]
=
\bigl(
Y^{\,int\,\sscr T}_{\sscr\cal L}\cdot
\alpha_{\sscr\cal L\cal L}\,
p^{\,ext}_{\sscr\cal L}
\bigr)-
\bigl(
Y^{\,int\,\sscr T}_{\sscr\cal L}\cdot
\alpha_{\sscr\cal L\cal L}
Y^{\,int}_{\sscr\cal L}
\bigr)=0.
\notag
\end{gather}
Two relations in the first line of Eqs.(5.13)
in our case of $\alpha$-parametric functions are analogs of
the first and the second Kirchhoff laws in electric networks.
Similarly, in the third line of Eqs.(5.12) we find in our case
an analog of the well-known expression for
a power dissipated in electric networks.

\vskip3mm

The present author wishes to express his gratitude thanks to Reviewer
for constructive review and valuable suggestions that enhances the improving
of the article.
The paper is based on the report presented
at the Bogolyubov Kyiv conference
``Modern Problems of Theoretical and Mathematical Physics'',
September 15-18, 2009, Kyiv, Ukraine.

\end{multicols}

\end{document}